\documentclass[3p,number,sort&compress]{elsarticle}
\usepackage{amssymb}
\usepackage{amsmath}
\usepackage{bm}
\usepackage[dvipsnames,table]{xcolor}
\usepackage{hyperref}
\hypersetup{
    pdftitle={Ab initio study on photocatalytic properties of PtSSe-WXY Janus heterostructures},    
    pdfauthor={},     
    pdfcreator={},   
    pdfproducer={}, 
    pdfkeywords={heterostructures} {water splitting} {photocatalysis} {janus transition metal dichalcogenides},
    colorlinks = true,
    citecolor  = ForestGreen,
    linkcolor  = RoyalBlue,
    urlcolor   = RoyalBlue
}
\usepackage{sectsty}
\usepackage{graphicx}
\usepackage{booktabs}
\usepackage{multirow}
\usepackage{array}
\usepackage{float}
\usepackage{fancyvrb}

\begin{document}

\begin{frontmatter}

\title{Ab initio study on photocatalytic properties of PtSSe-WXY Janus heterostructures}

\author[a]{Shivprasad S. Shastri \fnref{fn1}}
\ead{shastshi@fel.cvut.cz}
\author[a]{Antonio Cammarata \fnref{fn2}}
\ead{cammaant@fel.cvut.cz}
\author[a,b]{Tomas Polcar}
\affiliation[a]{organization={Department of Control Engineering, Faculty of Electrical Engineering,
Czech Technical University in Prague},
addressline={Karlovo Namesti 13},
city={Prague 2},
postcode={12135},
country={Czech Republic}}
\affiliation[b]{organization={Engineering Materials \& nCATS, FEE, University of Southampton},
city={Southampton},
postcode={SO17 1BJ},
country={United Kingdom}}
\fntext[fn1]{Corresponding author}
\fntext[fn2]{Corresponding author}

\begin{abstract}
Semiconductor photocatalysis offers a sustainable route for converting solar energy into chemical energy, enabling the production of clean fuels and valuable chemical products.
To this aim, we explore van der Waals heterostructures made up of Janus PtSSe and WXY (X, Y $=$ S, Se, Te and X $\neq$Y), in the context of photocatalytic applications.
The redox capabilities of various heterostructure configurations (atom facing types and stacking orders) are evaluated by aligning the absolute band edge positions with respect to redox potentials of hydrogen and oxygen evolution reaction (HER and OER) and CO$_2$ reduction reactions.
The stability of photocatalyst candidates are checked by layer binding energy calculations and ab initio molecular dynamics simulations.
The optical absorption spectra suggest good light absorption in the visible range.
Further, strain engineering is applied as a way to tune band edges and evaluate the possible use of the heterostructures as photocatalysts.
This study shows that van der Waals heterostructure bilayers composed of Janus PtSSe and WSeTe in specific geometric configurations can be potential materials as photocatalysts for HER, OER and CO$_2$ reduction reactions.
Finally, we suggest that, although systems made up of PtSSe and WSTe cannot be used for photocatalytic applications, they can be explored for applications in thermoelectric energy conversion or infrared photovoltaics. 
\end{abstract}

\end{frontmatter}

\addtocontents{toc}{\protect\setcounter{tocdepth}{0}}
\section{Introduction}
\label{sec:intro}
The conversion of solar energy into chemical energy through photocatalysis provides an alternative pathway for generating clean, renewable, and sustainable fuels and useful chemical products \cite{intro1,intro2,intro3,intro4}.
Photocatalytic reduction of CO$_{2}$ into useful chemicals such as alkanes, alcohols and aldehydes is one of the viable strategies to reduce CO$_{2}$ emissions or capturing it from the environment \cite{intro_acscat,intro_aksingh}.
Generation of hydrogen and oxygen gases using photocatalytic materials by splitting water promises environment friendly, clean and renewable energy source \cite{intro_hennig,intro_aksingh,intro_heroer}.
To this aim, development of such photocatalytic materials with tunable electronic properties, optimized for light absorption and band edge positions suitable for driving these desired chemical reactions, is crucial \cite{intro_aksingh2}.

Many materials have been explored for photocatalytic water splitting to produce hydrogen and oxygen, including 2D materials \cite{tio2,intro_2dpc1,intro_2dpc2,intro_2dpc3,intro_2dpc4,intro_2dpc5,intro_2dpc6,intro_2dpc7}.
Similarly, number of bulk and 2D materials are studied for photocatalytic reduction of CO$_2$ \cite{intro_acscat,intro_co2rr1,intro_co2rr2,intro_co2rr3}.   
2D materials have been actively studied for their use in catalysis, photovoltaics, battery electrodes, electronics, optoelectronics and their interesting new physics \cite{geim2d,intro_2dappln1,intro_2dappln2,intro_2dappln3}.
Relative to bulk materials, 2D materials are characterised by large surface area for photon absorption, enhanced charge separation, structural flexibility, cocatalyst integration and higher surface-water interaction;\cite{intro_hennig,2dadv1,2dadv2,2dadv3}.
these features are present thanks to their reduced thickness which shortens carrier migration distances to the surface, then reducing recombination possibility and providing more carriers for reactions \cite{intro_hennig,intro_2dpc7}.
In 2D monolayer semiconductors both electron and hole generation needed for redox reactions takes place in the same surface, resulting in possible high recombination of carriers \cite{2ddadv1,2dadv2}.
Arranging different 2D semiconductor materials in stackings to create heterostructures (HS) can be a promising strategy to design and tailor materials for specific properties, functionalities and to overcome the drawback of high carrier recombinations \cite{intro_2dpc7,novo_hs,hs_adv1,hs_adv2}.
These HSs can show type-I (straddled), type-II (staggered) or type-III (disjoint) band alignments \cite{intro_2dpc7}.
The type-I HS can be used for photocatalysis, although they can be less effective because of limited ability to separate the photogenerated charge carriers;
instead, type-II band alignment facilitates electrons and holes generation separately in two distinct stacked layers, reducing carrier recombination and showing enhanced performance \cite{intro_2dpc7,hs_adv2,hs_adv3,hs_adv4}.

2D materials such as transition metal monosulfides, graph-itic carbon nitrides, transition metal dichalcogenides (TMDC), ZnO, Blue P and their HSs are currently studied for photocatalytic applications \cite{intro_2dpc7,hs_adv2,tmdc_pc}.
In particular, the Janus type of TMDCs have attracted interest due to inherent electric field and dipole moment arising from the out of plane mirror symmetry breaking \cite{jtmdc_efield,jtmdc_pcreview}.
In contrast to TMDCs, the intrinsic dipole moment in Janus TMDCs has been reported to help in photo-induced electron-hole separation and elongating carrier recombination time, an important feature to collect charges and improve efficiency in photocatalytic materials \cite{jtmdc_efield,jtmdc_pcreview,jtmdchs_adv2}.
Janus TMDCs such as MoXY, WXY, PtSSe and HSs like MoX$_2$-MoXY, WX$_2$-WXY (X, Y $=$ O, S, Se, Te and X $\neq$Y) are explored considering the effect of intrinsic dipole moments on band edge alignment \cite{mosse,moxy,bilayermosse,ptssepc1,ptssepc2,wssepc1,wssepc2,hs_adv3}.
However, the HSs made up of PtSSe and WXY monolayers have not yet been studied for photocatalytic water splitting or CO$_2$ reduction reactions.
This motivates the present study, which aims to contribute to the discovery of new Janus heterostructures for photocatalysis applications.

In this work, we consider TMDC-based Janus heterostructures, namely 1T-PtSSe/2H-WSSe (PtWSSe), 1T-PtSSe/2H-WSTe (PtWSTe) and 1T-PtSSe/2H-WSeTe (PtWSeTe), and study the stability, electronic and optical properties at various interface configurations and stacking orders.
By checking suitable alignment of the band edge positions with respect to vacuum level and comparing with oxidation-reduction potentials, we suggest that these materials are good candidates for hydrogen evolution reaction (HER), oxygen evolution reaction (OER) and CO$_2$ reduction applications.
\section{Computational Methods}
\label{sec:methods}

\begin{figure*}[ht]
    \centering
     \includegraphics[width=16cm,height=7cm]{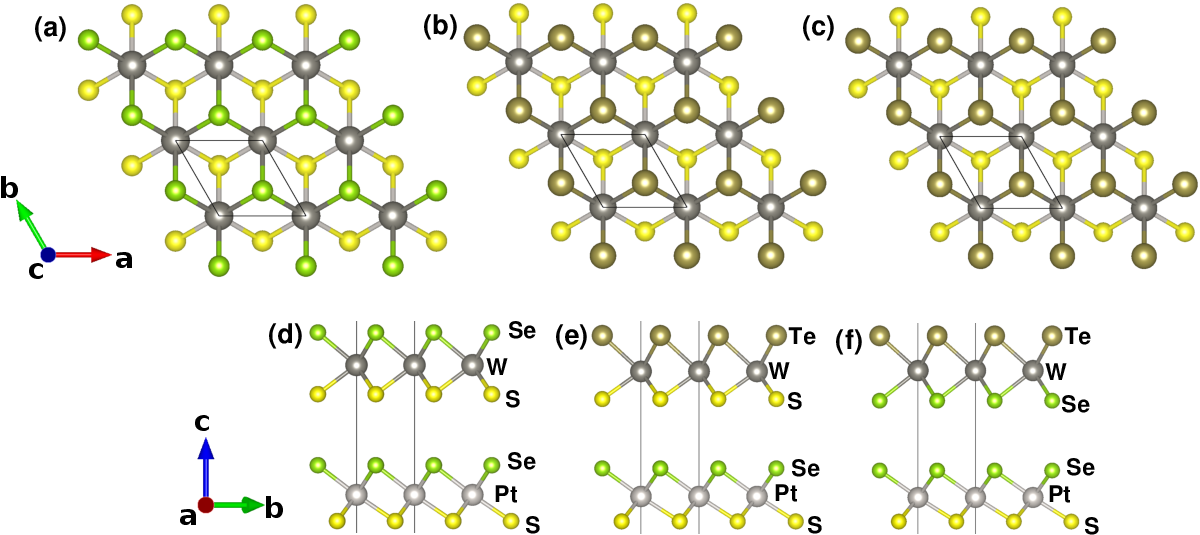}
     \caption{Model geometry of three HSs considered in this work: (a) PtSSe-WSSe (b) PtSSe-WSTe and (c) PtSSe-WSeTe (top view). Fig. (d)-(f) show the respective side views.}
     \label{fig:hsstruct}
\end{figure*}

We choose 1T-PtSSe and 2H-WXY (X, Y $=$ S, Se, Te and X $\neq$Y) 2D monolayers to create bilayer heterostructures.
Previous ab initio studies show that these monolayers display photocatalytic HER, OER and water splitting properties \cite{ptssepc1,ptssepc2,wssepc1,wssepc2}.
The reported cohesive energy values for these monolayers suggests that  their formation is energetically favoured \cite{ptssepc2,wxy_formphonon}, while related phonon spectra support their dynamical stability \cite{ptssepc1,ptssepc2,wxy_formphonon}.
Thus, we consider these monolayers to build three vertically stacked bilayer HSs in the following way.
In our geometric models, we position the monolayers parallel to the crystallographic ($\bm{a}$, $\bm{b}$) plane;
we then stack the 1T-PtSSe monolayer as the lower layer and 2H-WXY as the upper layer along the $\mathbf{c}$ crystallographic axis orthogonal to the monolayers plane.
The initial in-plane lattice parameters are chosen to be the average of the optimized lattice parameters of the parent monolayers.
We then perform full (both lattice and atom position) structure optimizations to get optimized lattice constants and atomic positions for three HSs;
the relaxed geometries used in this work are reported in the last section of the Supplementary Information (SI).
We use the following notation to refer to the HSs models:
\textit{i)} PtWSSe  (PtSSe-WSSe),
\textit{ii)} PtWSTe (PtSSe-WSTe),
and \textit{iii)} PtWSeTe (PtSSe-WSeTe).
The top and side views of these model structures are shown in \autoref{fig:hsstruct}.

\begin{figure}[h]
    \centering
     \includegraphics[height=7.5cm,width=9cm]{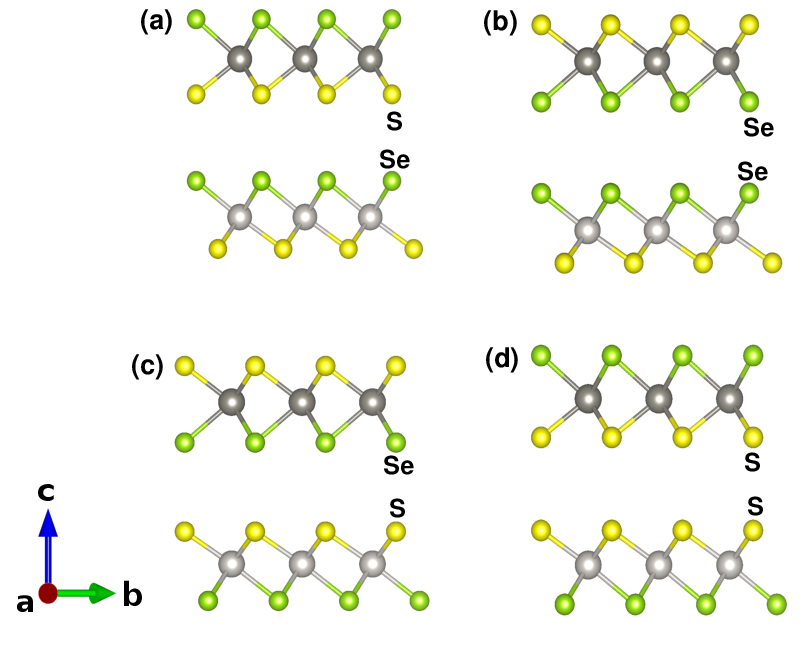}
     \caption{The four possible atom facing configurations, IC1-IC4 ((a)-(d)) in the case of PtWSSe HS. Similarly, four possible ICs are considered in case of other two HSs.}
     \label{fig:ic_wsse}
\end{figure}

Since the parent Janus monolayers shows a lack of inversion symmetry, different atom-facing combinations can be considered at the heterostructure interface.
The type of atoms facings at the interface can lead to changes in the electronic structure and band gap due to the overlap of different atom pairs, then acting as a possible way to tune the electronic properties.
For a given HS, four atom facing configurations (or interface configurations) are possible at the interface \textit{viz.} S-X , S-Y, Se-X, and Se-Y for a fixed X and Y atom types of the top layer (WXY layer);
we will refer to them as IC1-IC4. 
An example of these four ICs (IC1-IC4) in the case of the PtWSSe system is shown in \autoref{fig:ic_wsse}.
Similarly, four ICs are considered in the case of PtWSTe and PtWSeTe HSs.
In the construction of a bilayer, different stacking orders are possible depending on the relative atomic positions of the top and bottom monolayers.
We consider five stacking orders which are denoted as S1-S5 in what follows;
related discussion is given in section ``Structural Properties'' of the SI, together with the five different stacking orders.
Both the types of interface configurations and stacking orders can affect the electronic structure and might serve to tune the band gap of the overall system.

We perform ab initio calculations within the framework of the Density Functional Theory (DFT) with the projector augmented wave method as implemented in the \textsc{vasp} software \cite{vasp1, vasp2}.
Structural relaxation (atomic position and lattice parameters) is carried out using Perdew-Burke-Ernzerhof (PBE) energy functional; \cite{pbe}
the DFT-D3 functional of Grimme et al. \cite{dftd3} is used to account for van der Waals interactions.
This choice of energy functional and van der Waals correction is motivated by their use for similar van der Waals materials \cite{dftd3_1,dftd3_2,dftd3_3,dftd3_4,dftd3_5}.
The value of the lattice constant $c$ is set to 30 {\AA} to avoid interactions between the images along the same direction.
The plane wave basis set is truncated with an energy cutoff equal to 500 eV, while the Brillouin zone is sampled using a 13$\times 13\times1$ Monkhorst-Pack mesh \cite{monkhorst76}.
The Self-Consistent Field and force convergence criteria are set to 10$^{-9}$ eV and 10$^{-4}$ eV \AA$^{-1}$, respectively. 
We perform hybrid DFT calculations to get more accurate band gap description using the Heyd-Scuseria-Ernzerhof (HSE06) functional \cite{hse06}.
These calculations are performed for those HS configurations which showed band gap more than 0.5 $\mathrm{eV}$ from preliminary PBE functional estimations, to exclude candidates with narrow band gaps.
The thermal stability of heterobilayers is checked by ab initio molecular dynamics (AIMD) simulations at 300 K using the NVT ensemble and Nos\'e-Hoover thermostat, using supercells with dimension 3$\times$3$\times$1 and a time step of 1 fs.
The frequency dependent dielectric function are obtained using the method of Gajdo\v{s} et al. as implemented in \textsc{vasp} \cite{loptics}
We post-process the data with the \textsc{vaspkit} utility to obtain optical absorption spectra, layer projected band structures and planar averaged electrostatic potentials \cite{vaspkit}.

\section{Results and Discussion}
\label{sec:results}
\subsection{Structural properties and stability}
\label{subsec:struc_stab}
The optimized lattice constants and the interlayer distances $d$ for stacking order S1 for all the ICs are reported in \autoref{tab:lat_ldist}, where $d$ is the distance between the layers of chalcogen atoms at the interface of the heterostructure. 
Analogous tables are reported in the SI for the stacking orders S2-S5 (Tables S1-S4).
\begin{table*}[t]
  \centering
  \caption{Optimized lattice constants ($a = b$) and interlayer distance $d$ [\AA{}] for three HSs in different interface configurations for stacking order S1.}
  \label{tab:lat_ldist}
  \resizebox{\textwidth}{!}{%
  \begin{tabular}{c c c c c c c}
    \toprule
    \multirow{2}{*}{Atom facing types} & 
    \multicolumn{2}{c}{PtWSSe} & 
    \multicolumn{2}{c}{PtWSTe} & 
    \multicolumn{2}{c}{PtWSeTe} \\
    \cmidrule(lr){2-3} \cmidrule(lr){4-5} \cmidrule(lr){6-7}
     & Lattice constant & $d$ & Lattice constant & $d$  & Lattice constant & $d$ \\
    \midrule
    IC1  & 3.4063 & 3.62 (Se-S) & 3.4773 & 3.60 (Se-S) & 3.5169 & 3.69 (Se-Se) \\
    IC2  & 3.4061 & 3.72 (Se-Se) & 3.4763 & 3.91 (Se-Te) & 3.5160 & 3.90 (Se-Te) \\
    IC3  & 3.4069 & 3.65 (S-Se) & 3.4773 & 3.80 (S-Te) & 3.5167 & 3.79 (S-Te) \\
    IC4  & 3.4071 & 3.57 (S-S) & 3.4779 & 3.55 (S-S) & 3.5175 & 3.62 (S-Se) \\
    \bottomrule
  \end{tabular}}
\end{table*}
By inspecting the table, we observe that changing the atom types at the interface has a smaller effect on the lattice constants, while it has a greater effect on the interlayer distance because of the different sizes of the atoms at the interface. 
\begin{table*}[ht]
  \centering
   \caption{Layer binding energy $E_b$ [$\mathrm{eV}$] for PtWSSe and PtWSeTe HSs with different ICs and stacking orders.}
  \label{tab:Eb}
  \resizebox{\textwidth}{!}{%
  \begin{tabular}{c c c c c c c c c }
    \toprule
    \multirow{2}{*}{Stacking order} &
    \multicolumn{4}{c}{PtWSSe} &
    \multicolumn{4}{c}{PtWSeTe} \\
    \cmidrule(lr){2-5} \cmidrule(lr){6-9} 
      & IC1 (Se-S) & IC2 (Se-Se) & IC3 (S-Se) & IC4 (S-S) & IC1 (Se-Se) & IC2 (Se-Te) & IC3 (S-Te) & IC4 (S-Se) \\ 
    \midrule
    S1  & 0.338 & 0.336  & 0.342 & 0.353 & -0.059 & -0.058 & -0.058 & -0.049 \\
    S2  & 0.261 & 0.259  & 0.246 & 0.263 & -0.162 & -0.152 & -0.182 & -0.167 \\
    S3  & 0.252 & 0.254  & 0.255 & 0.265 & -0.161 & -0.150 & -0.161 & -0.154 \\
    S4  & 0.260 & 0.255  & 0.256 & 0.274 & -0.153 & -0.157 & -0.168 & -0.144 \\
    S5  & 0.268 & 0.268  & 0.262 & 0.268 & -0.150 & -0.128 & -0.153 & -0.160 \\
    \bottomrule
  \end{tabular}}
 
\end{table*}
The stability of the HSs is analysed by considering the layer binding energy $E_b$ and the results of AIMD simulations at 300 K.
The layer binding energy $E_b$ is defined as the difference between the energy of the heterostructure $E_{HS}$ and the sum of the energy of the constituent monolayers $E_{ML1,2}$:
$E_{b} = E_{HS} - (E_{ML1} + E_{ML2})$.
The calculated $E_b$ for PtWSSe and PtWSeTe HS are tabulated in \autoref{tab:Eb}, while for the PtWSTe systems the $E_b$ values are given in Table S5 of the SI.
The constituent monolayers already show synthesis feasibility by cohesive energy calculations\cite{ptssepc2,wxy_formphonon}, and were found to be dynamically stable by phonon dispersion calculations \cite{ptssepc1,ptssepc2,wxy_phonon,wxy_formphonon}
The calculated layer binding energy values for PtWSTe and PtWSeTe HSs are negative, which supports the energetic feasibility of bilayer formation.
In the case of PtWSSe HSs, the values of $E_b$ are positive, suggesting that layer binding is less favourable and can be obtained under suitable synthesis conditions;
for this reason, we do not further consider the PtWSSe HSs in what follows.

\begin{figure*}[ht]
    \centering
     \includegraphics[width=0.85\textwidth]{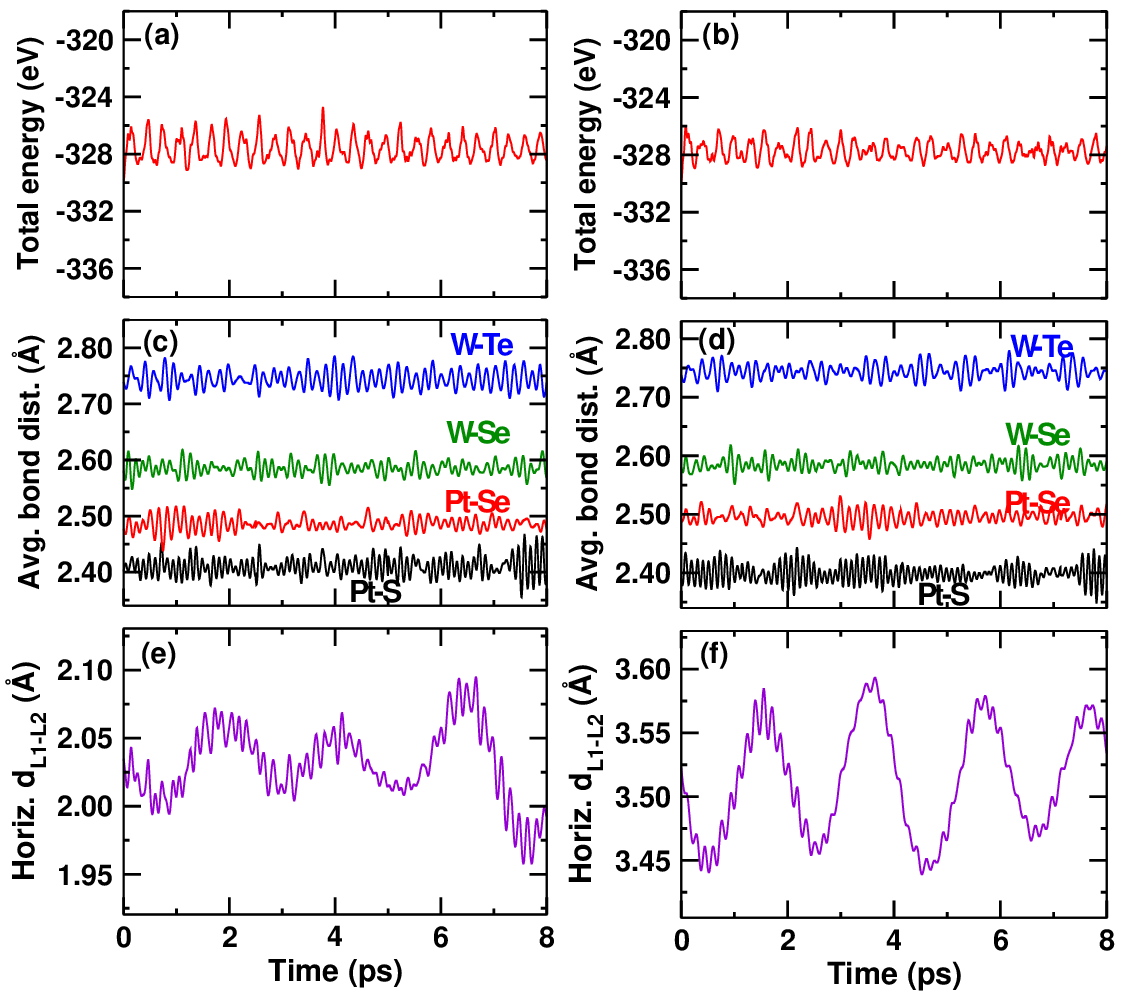}
     \caption{Variation of total energy of the system, average bond distance (Avg. bond dist.) and horizontal distance between layers (Horizon. d$_{\mathrm{L1-L2}}$) as a function of time. Left and right panels show the results for PtWSeTe-IC1-S2 and PtWSeTe-IC4-S2 systems, respectively.}
     \label{fig:aimd1}
\end{figure*}

Results of AIMD simulations at 300 K are shown in \autoref{fig:aimd1}.
The variation of the total energy of PtWSeTe-IC1-S2 and PtWSeTe-IC4-S2 HSs as function of time is shown in \autoref{fig:aimd1}a and \autoref{fig:aimd1}b, respectively.
To check for possible breaking of the atomic bonds, we monitored the average bond distance between the metal and first neighbouring chalcogenide atoms over the simulation time (\autoref{fig:aimd1}c and d).
For both HSs, less than 0.1 \AA{} variation in the bond length is observed.
In addition, the possibility of relative layer sliding compared to the initial stacking order is checked (\autoref{fig:aimd1}e and f).
To this aim, we monitor the horizontal distance between the metal atoms of upper (WXY) and lower (PtSSe) layers.
The horizontal distance at a given time step is calculated from the average of $a$ and $b$ coordinates of upper and lower layer metal atoms.
The plots show a change of less than 0.1 \AA{} in the horizontal distance from the initial position, which indicates that the initial stacking order is stable;
similar plots for PtWSeTe-IC2-S2 HS are shown in Figure S3 of the SI.
These results suggest that the HSs are stable at operating temperature.

\subsection{Electronic properties}
\label{subsec:elecprop}
\begin{figure*}[h]
    \centering
     \includegraphics[width=0.7\textwidth]{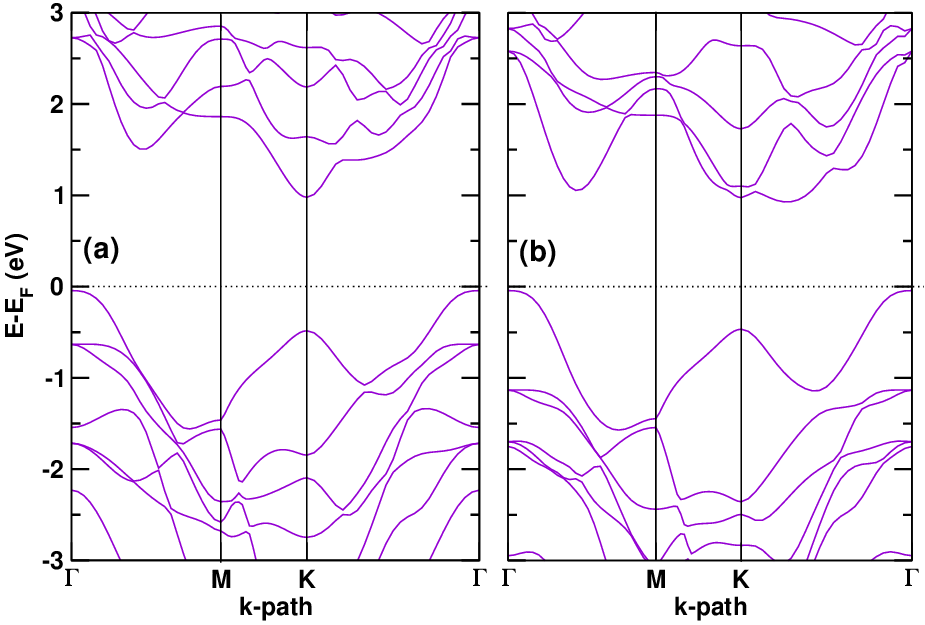}
     \caption{Electronic band structure of (a) PtWSeTe-IC1-S2 and (b) PtWSeTe-IC4-S2 heterostructures calculated using the HSE06 functional.}
     \label{fig:estruct1}
\end{figure*}

In this section, we discuss the electronic structure of selected candidate HS configurations (ICs and stacking orders) which can show photocatalytic activity.
We individuate the type of band gap, value of band gap and the band edge positions in these HSs, which are useful features for photocatalytic redox reactions. 
\autoref{fig:estruct1} shows the electronic structure of PtWSeTe-IC1-S2 and PtWSeTe-IC4-S2 HSs obtained using HSE06 functional.
The PtWSeTe HS has Se-Se atom facing in IC1 and S-Se atom facing in IC4 configurations at the interface;
both HSs are indirect band gap semiconductors.
The values of the band gap $E_g$ for PtWSeTe in IC1-S2 and IC4-S2 configurations are 1.02 eV and 0.97 $\mathrm{eV}$, respectively (\autoref{tab:band_edge}).
For both the HSs, the valence band maximum (VBM) is located at the $\Gamma$-point. 
The conduction band minimum (CBM) is realised at the $K$-point for IC1-S2 while it is slightly away from the $K$-point along $\Gamma$-$K$ direction (closer to the $K$-point) for IC4-S2 configuration.
In the case of PtWSeTe-IC4-S2, two other minima are observed:
one is located at the $K$-point, close to the CBM with an energy difference of less than $\sim$10 meV and another along the $\Gamma$-$M$ path realising a band gap of 1.05 $\mathrm{eV}$.
These minima are also expected to contribute to the carrier (electron-hole pair) generation during photon absorption.
With the availability of multiple minima with close energy values, high carrier generation is expected in this HS.
The direct gap widths in case of PtWSeTe for IC1-S2 and IC4-S2 configurations are $\sim$1.47 and $\sim$1.43 $\mathrm{eV}$, respectively.
\begin{figure*}[ht]
    \centering
     \includegraphics[width=0.7\textwidth]{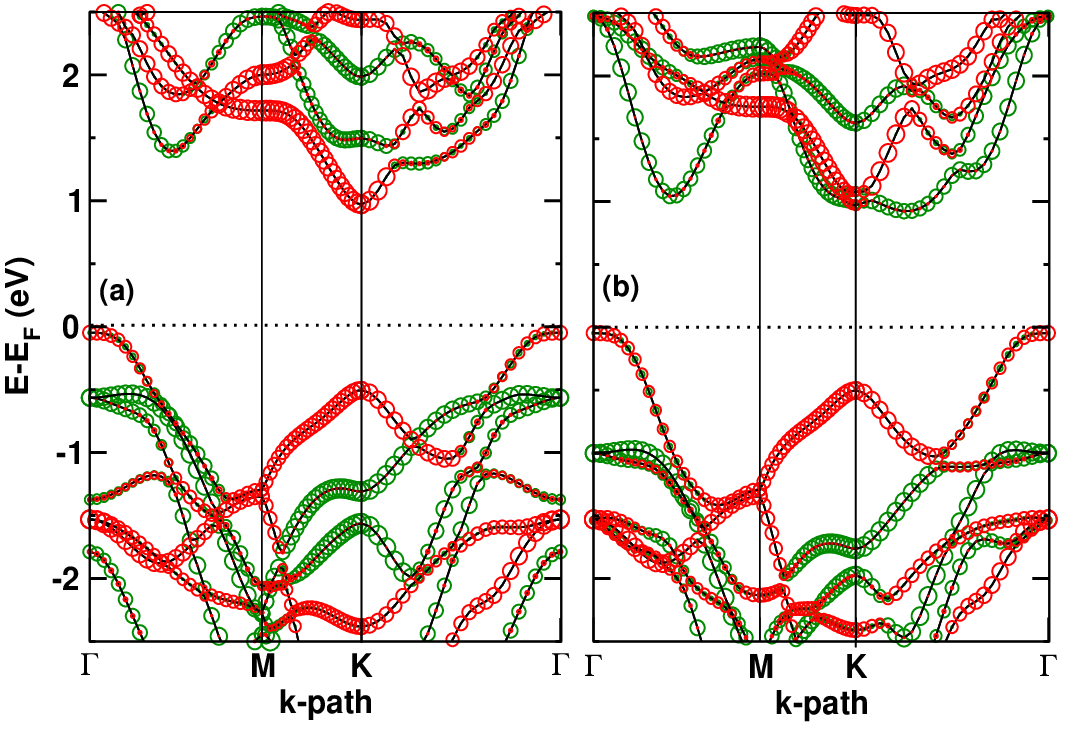}
     \caption{Layer projected band structures of (a) PtWSeTe-IC1-S2 (b) PtWSeTe-IC4-S2 systems. Red and green circles represent projection onto WSeTe and PtSSe component monolayers, respectively; larger circles indicate larger component contributions.}
     \label{fig:lproj}
\end{figure*}
To check the location of band edges (CBM and VBM) and verify the type of band alignment, we consider the atom-projected band structures onto component monolayer character.
The layer character of an electronic band is obtained as the sum of the atom-projected character of each atom forming the layer. 
From \autoref{fig:lproj}a, we can see that VBM ($\Gamma$-point) and CBM ($K$-point) own WSeTe character, suggesting straddled type of alignment;
thus, the PtWSeTe system with atom facing type IC2 and stacking S2 has type-I band alignment.
For PtWSeTe (\autoref{fig:lproj}b), it can be observed that the VBM has WSeTe character while CBM has PtSSe character, suggesting this HS in IC4-S2 configuration is of type-II (staggered) band alignment.
The presence of VBM and CBM in two separate layers is beneficial to generate carriers in two separate layers and is then helpful in reducing carrier recombination.
Also, here we note that changing the atom facing type is a useful way to tune the band alignment type between type-I and type-II.

Lastly, we here provide a brief discussion on the electronic structure of PtWSTe HSs.
On changing the atom facing type and stacking orders, the PtWSTe HSs show electronic band structures ranging from metallic to semiconducting with the highest band gap value equal to 0.84 eV.
The electronic band structure for a case of PtWSTe in IC1 atom facing type for the most stable stacking order S3 is shown in Figure S4 of SI.  
The VBM and CBM are realised at the $\Gamma$- and $K$-point, respectively, while the band gap is found to be indirect with a width equal to 0.57 eV.
We do not further explore and discuss these HSs for photocatalytic applications as their band gap are lower for efficient absorption of visible light;
however, these HSs can be explored for other applications such as thermoelectric energy conversion and infrared photovoltaics where narrow band gap semiconductors are useful \cite{thermoelectric,ir_phvlt1,ir_phvlt2,ir_phvlt3,ir_phvlt4,ir_2d}.
\subsection{Photocatalytic activity}
\label{subsec:photocat}

\begin{figure}[ht]
    \centering
     \includegraphics[width=0.45\textwidth]{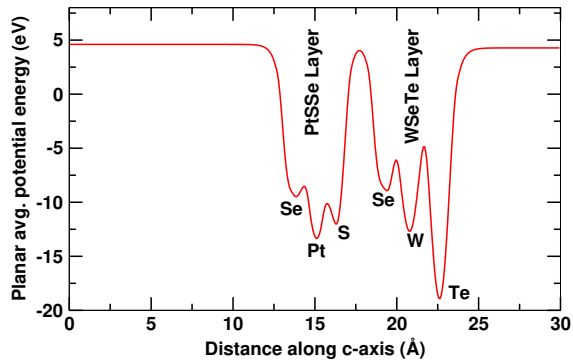}
     \caption{Planar averaged electrostatic potential along $\mathbf{c}$-axis for PtWSeTe-IC4-S2. }
     \label{fig:pl_avg}
\end{figure}
In this section, we discuss the band edge alignments in the selected HSs for HER, OER and CO$_2$ reduction applications.
The difference in electronegativity of the chalcogen atoms in the considered Janus HSs generates a dipole moment and an internal electric field. 
Moreover, because of the electronegativity difference, the two sides (upper WXY and lower PtSSe sides --- \autoref{fig:hsstruct}) of the HS bilayer have different vacuum levels.
In the cases where the VBM and CBM are located in two different layers, we align the VBM and CBM edges with respect to the vacuum levels of the respective sides following the method by X. Li \textit{et al.} \cite{prl_align}.
 We present the $ab$-planar averaged electrostatic potential along $\mathbf{c}$-direction for PtWSeTe-IC4-S2 as a case, which possesses a type-II band gap, in order to show the vacuum potential differences arising due to the chalcogen atoms of different electronegativity in the considered HSs.
From \autoref{fig:pl_avg}, difference in vacuum potentials at the PtSSe side (4.60 eV) and at the WSeTe side (4.27 eV) and the asymmetric nature of the potential peaks can be observed. 
This asymmetric nature of the potential curves indicates an internal electric field arising due to the presence of chalcogen atoms of different electronegativity.

The hydrogen and oxygen evolution reactions occur thanks to the following reduction and oxidization processes:
\begin{align}
\mathrm{2H^+ + 2e^-} &\rightarrow \mathrm{H_2} \label{eq:her}\\
\mathrm{H_2O + 2h^+} &\rightarrow \tfrac{1}{2}\mathrm{O_2 + 2H^+}. \label{eq:oer}
\end{align}
The reduction and oxidation potential of the above reactions at pH=0 are -4.44 eV and -5.67 eV, respectively, with respect to the vacuum level  \cite{intro_hennig}.
Similarly, the photoreduction of CO$_{2}$ to HCOOH, CO and other molecules takes place at pH=7 through the following multielectron steps \cite{co2red1,co2red2}:
\begin{align}
\mathrm{CO_2 + 2H^+ + 2e^-} &\rightarrow \mathrm{HCOOH}  \label{eq:co2_1} \\
\mathrm{CO_2 + 2H^+ + 2e^-} &\rightarrow \mathrm{CO + H_2O}  \label{eq:co2_2} \\
\mathrm{CO_2 + 4H^+ + 4e^-} &\rightarrow \mathrm{HCHO + H_2O}  \label{eq:co2_3} \\
\mathrm{CO_2 + 6H^+ + 6e^-} &\rightarrow \mathrm{CH_3OH + H_2O}  \label{eq:co2_4} \\
\mathrm{CO_2 + 8H^+ + 8e^-} &\rightarrow \mathrm{CH_4 + 2H_2O}.  \label{eq:co2_5}
\end{align}
The reduction potentials for reactions in Equations (\ref{eq:co2_1})-(\ref{eq:co2_5}) with respect to vacuum level are -3.89, -3.97, -4.02, -4.12 and -4.26 eV, respectively, which are obtained using the relation in Ref. \cite{echem_scale_conv}.
For reduction reaction to take place, the CBM edge should be higher in energy than the reduction potential;
similarly for oxidation to take place the VBM edge should be lower in energy than the oxidation potential.
To assess the photocatalytic redox activities of the selected HSs, we then focus on the CBM and VBM edge positions aligned with respect to the vacuum level.
\autoref{tab:band_edge} shows candidates which can act as photocatalysts for HER and CO$_2$ reduction reactions. %
\begin{table}[ht]
\caption{Band gap, band alignment type, vacuum level, valence band edge E$_{VBM}$ and conduction band edge E$_{CBM}$ with respect to vacuum level for HS candidates. All the energies are in $\mathrm{eV}$.}
\label{tab:band_edge}
\centering
\resizebox{\textwidth}{!}{%
\begin{tabular}{lcccccc}
\hline\hline
  & 
E$_{g}$ & 
Band alignment &
Vacuum level (PtSSe side) & 
Vacuum level (WSeTe side) & 
E$_{VBM}$ & 
E$_{CBM}$  \\
\hline
PtWSeTe-IC1-S2  & 1.02 & type-I & 5.27 & 3.63 & -4.35 & -3.33  \\
PtWSeTe-IC4-S2  & 0.97 & type-II & 4.60 & 4.27 & -4.54 & -3.89  \\
\hline
\end{tabular}}

\end{table}
In the table, the valence (E$_{VBM}$) and conduction (E$_{CBM}$) band edge energies are obtained by aligning them with respect to the vacuum level. 
In case of candidates with band edges located in the same layer, the band edges are aligned with respect to the vacuum level of the component monolayer side in which both edges (VBM and CBM) are located. 
By inspecting the table, we can see that both the PtWSeTe-IC1-S2 and PtWSeTe-IC4-S2 HS can act as photocatalytsts for HER with E$_{CBM}$ values higher than water reduction potential of -4.44 $\mathrm{eV}$.
The type II nature of PtWSeTe-IC4-S2 can be an advantage for efficient carrier generation in separate layers for HER with multiple minima contributing to higher number of conduction band electrons.
Instead, the efficiency of PtWSeTe-IC1-S2 can be lower because of the type-I nature of the alignment, favouring possible carrier recombination of electrons-hole pairs generated in the same layer.
We find that these HSs do not show OER with E$_{VBM}$ edges higher in energy than the water oxidation potential of -5.67 $\mathrm{eV}$.

\begin{figure*}[t]
    \centering
     \includegraphics[width=16.2cm,height=7cm]{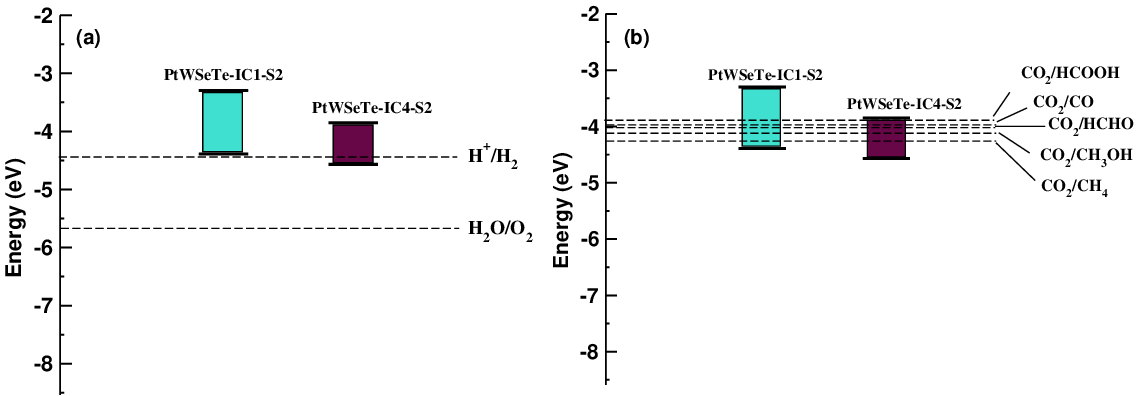}
     \caption{Alignment of band edge positions of different heterostructures and redox potentials relative to the vacuum level. (a) Alignment of band edge positions and water redox potential and (b) alignment of band edge positions and CO$_2$ reduction potentials.}
     \label{fig:redox}
\end{figure*}
Concerning the CO$_2$ reduction capabilities, by comparing the potentials of multielectron reduction of CO$_{2}$ in aqueous medium (Equations (\ref{eq:co2_1})-(\ref{eq:co2_5})) and the CBM energies from \autoref{tab:band_edge}, we find that PtWSeTe-IC1-S2 can be a suitable candidate.
With its conduction band edge above these reduction potential, it can act as photocatalysts for $\mathrm{CO}$, $\mathrm{HCHO}$, $\mathrm{CH_{3}OH}$ and $\mathrm{CH_{4}}$ conversion of CO$_{2}$.
The other candidate PtWSeTe-IC4-S2 has also the suitable conduction band edge to drive the reduction reactions of CO$_2$ except for the $\mathrm{HCOOH}$ conversion.
However, for some of the above reactions, the overpotentials (the energy difference between the reduction (oxidation) potential and CBM (VBM)) to drive the reactions can be low, requiring the need to supply additional energy by an external electric potential difference.
Such materials can then be suitable candidates for photoelectrocatalysts in HER and CO$_2$ reduction reactions.

To get a better picture of the band edge alignments we provide energy level diagrams in which we report the alignment of the band edge positions and redox potentials with respect to the vacuum level (\autoref{fig:redox}).
HER and OER activities can be estimated from \autoref{fig:redox}a, while CO$_{2}$ reduction activities from \autoref{fig:redox}b.
It can be observed that the CBM position of PtWSeTe IC1-S2 and PtWSeTe IC4-S2 favours the donation of electrons to H$^{+}$ ions for the reduction reaction.
\autoref{fig:redox}b suggests that PtWSeTe IC1-S2 and PtWSeTe IC4-S2 HSs have conduction band edge positions favourable to drive the CO$_2$ reduction reaction to useful molecules.
Moderate values of overpotentials ($\sim$0.5-1.0 $\mathrm{eV}$) can be observed for the two candidates to drive these reactions, except for the case of CO$_2$ reduction with PtWSeTe-IC4-S2.

\begin{figure}[h]
    \centering
     \includegraphics[width=0.4\textwidth]{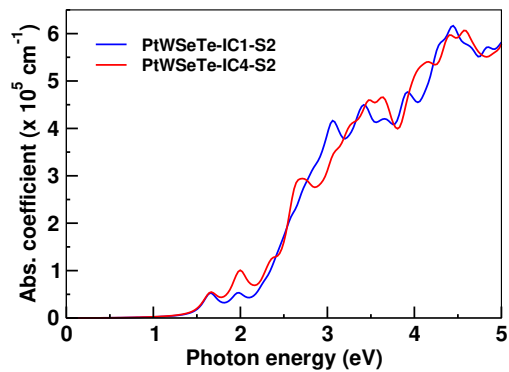}
     \caption{Calculated optical absorption coefficient $\alpha$ for PtWSeTe-IC1-S2 and PtWSeTe-IC4-S2 heterostructures. }
     \label{fig:alpha}
\end{figure}
In addition to possessing a favourable band edge alignment, a photocatalytic material needs to show a good incident light absorption property for overall efficiency.
Therefore, we calculate the optical absorption coefficient $\alpha$ as a function of incident photon energy for the HSs which show favourable band edge alignments for redox reactions, namely the PtWSeTe-IC1-S2 and PtWSeTe-IC4-S2 systems (\autoref{fig:alpha}). 
$\alpha$ is obtained by using the real and imaginary parts of the complex dielectric function from the HSE06 calculations.
The spectra show absorption peaks in the visible region because of direct transitions for both candidates.
The PtWSeTe-IC1-S2 system shows higher absorption of incident photons around 2 $\mathrm{eV}$ while, PtWSeTe-IC4-S2 shows higher values of the absorption coefficient for incident photon energy around 3 $\mathrm{eV}$.
The calculated absorption spectra suggest these HSs can utilise visible part of the solar spectrum by direct transitions then drive photocatalytic redox reactions in an efficient way.

\subsection{Band edge tuning by strain}
\label{subsec:edgetune}
Application of strains is one of the techniques used to engineer the positions of band edges of materials \cite{intro3,strain_hennig}
Here, we further study the effect of biaxial strain on the band edges of the HSs and check if they can be made suitable for photocatalytic applications by tuning the band edge positions with respect to redox potentials. 
We apply biaxial compressive strain in the range $[-4, -1] \%$ and tensile strain in the range $[+1, +4] \%$ of the initial lattice constants, and calculate the related electronic band structure and vacuum potentials using the HSE06 functional.
Here, we present and discuss only those HS configurations that showed suitable band edge alignment for photocatalytic activity, namely PtWSeTe-IC1-S2 and PtWSeTe-IC2-S2.

The general variation of band gap for the full range of applied strains for these HSs is reported in \autoref{tab:strain_gap}.
The application of tensile strain reduces the band gap width in both systems, while compressive strain increases the gap width for PtWSeTe-IC1-S2, for strain up to -3.0 $\%$ and upto -1.0 $\%$ for PtWSeTe-IC2-S2.
\begin{table}[h]
\centering
\begin{tabular}{ccc}
\hline\hline
Strain ($\%$)  & PtWSeTe-IC1-S2  & PtWSeTe-IC2-S2 \\
\hline
-4.0 & 1.45 & 0.99   \\
-3.0 & 1.50 & 1.07   \\
-2.0 & 1.46 & 1.14   \\
-1.0 & 1.24 & 1.22   \\
0.0 & 1.02 & 1.21   \\
1.0 & 0.82 & 1.18   \\
2.0 & 0.65 & 1.00   \\
3.0 & 0.48 & 0.82   \\
4.0 & 0.35 & 0.67   \\
\hline
\end{tabular}
\caption{Band gap [eV] of PtWSeTe-IC1-S2 and PtWSeTe-IC2-S2 systems at the considered compressive and tensile strain values.}
\label{tab:strain_gap}
\end{table}
\begin{table*}[h]
\centering
\begin{tabular}{lccccccc}
\hline\hline
  & Strain & $E_g$ &  Band alignment & Vacuum level  & Vacuum level  &  E$_{VBM}$ & E$_{CBM}$  \\
 & (\%) & &  & (PtSSe side) &(WXY side) & & \\
\hline
PtWSeTe-IC1-S2 & -1.0 & 1.24 & type-I & 5.34 & 3.71 & -4.41 & -3.17  \\
PtWSeTe-IC2-S2 & -1.0 & 1.22 & type-II & 4.92 & 4.76 & -5.72 & -4.66  \\
               & -2.0 & 1.14 & type-II & 4.99 & 4.86 & -5.76 & -4.74  \\
\hline
\end{tabular}
\caption{Strain, band gap, band alignment, vacuum level, valence band edge energy E$_{VBM}$ and conduction band edge energy E$_{CBM}$ with respect to vacuum level for HS candidates. All the energies are in $\mathrm{eV}$.}
\label{tab:strain_table}
\end{table*}
The values of band gaps, type of band alignment, vacuum level, band edge positions with respect to vacuum, particularly for those strain percentages which adjust the band edge positions for redox reactions are reported in \autoref{tab:strain_table}.  
Before discussing the photocatalytic activity of these strained HSs, we give a brief discussion of electronic structure and layer projected band characters of them.
The electronic band structure of pristine (unstrained) PtWSeTe-IC2-S2 in shown in Figure S5 of SI and that of PtWSeTe-IC1-S2 is shown in \autoref{fig:estruct1}a. 
The band structures and type of band alignment for the strained HSs are discussed in section ``Band edge tuning by strain'' of SI.
We find that PtWSeTe-IC2-S2 is of type-II at -1.0 and -2.0 $\%$ strain, while PtWSeTe-IC1-S2 is of type-I at -1.0 $\%$ strain.

Comparison of hydrogen reduction potential and the conduction band edges suggests that PtWSeTe-IC1-S2 can be a potential candidate for HER thanks to a higher E$_{CBM}$ compared to the potential for the reaction;
however, the efficiency of the reaction might be limited due to the type-I nature of the band alignment.
For OER, the comparison of E$_{VBM}$ with the oxidation potential of water (-5.67 $\mathrm{eV}$) suggests that strained PtWSeTe-IC2-S2 HSs (-1.0 and -2.0 $\%$ strain) can be potential candidates with better efficiency thanks to the type-II nature of the band alignment.
Finally, the CO$_{2}$ reduction reaction is favourable only by the conduction band edge position of strained PtWSeTe-IC1-S2 HS, although the efficiency might be low due to the type-I nature of the band alignment in this material.
The relative alignment of band edge positions of these strained HSs and redox potentials are graphically represented in Figure S8 of SI.
We observe that valence band edges of strained PtWSeTe-IC2-S2 can accept electrons from water molecules, acting as promising candidate for OER. 
Finally, we find that the conduction band edge of strained PtWSeTe-IC1-S2 has favourable position to donate electrons for HER and CO$_{2}$ reduction reaction.

\section{Summary}
\label{sec:summary}
In this work, we study three bilayer heterostructures formed by one Janus PtSSe and one WXY monolayer, namely PtWSSe, PtWSTe and PtWSeTe, in the context of photocatalytic applications. 
Four atom facing types and five different stacking orders are considered.
The suitability of the HS configurations for photocatalytic redox reactions is evaluated by aligning the ab initio calculated band edges with respect to redox potentials of HER, OER and CO$_2$ reduction reactions. 
We find that PtWSeTe-IC1-S2 and PtWSeTe-IC4-S2 HSs can be promising candidates for HER and CO$_2$ reduction reactions, whereas the type-II band alignment of PtWSeTe-IC4-S2 suggests higher reaction efficiency in this material.
Optical absorption spectra indicate that both systems exhibit good visible-light absorption.
The stability of these candidate HSs is supported by formation energy values and AIMD simulations at 300 K.
In addition, biaxial strain engineering is used to tune the band edge positions, in order to favour redox reactions.
Under compressive strain, PtWSeTe-IC1-S2 (-1.0 $\%$) emerges as a potential candidate for HER and CO$_{2}$ reduction, while PtWSeTe-IC2-S2 (-1.0 $\%$ and -2.0 $\%$) can act as a photocatalyst for OER.
These results indicate that HSs composed of Janus PtSSe and WSeTe monolayers are promising materials to be explored for photocatalytic applications.
Finally, we suggest that PtWSTe HSs with narrow band gaps can be explored for use in  thermoelectric generator or infrared photovoltaics applications.

\section*{Conflicts of interest}
There are no conflicts to declare.

\section*{Acknowledgements}
\label{sec:acknow}
This work was supported by the CTU Global Postdoc Fellowship programme.\\
This work was co-funded by the European Union under the project ``Robotics and advanced industrial production'' (reg. no. CZ.02.01.01/00/22\_008/0004590).
This work was supported by the Ministry of Education, Youth and Sports of the Czech Republic through the e-INFRA CZ (ID:90254).
Use of \textsc{vesta} software \cite{vesta} is also acknowledged.

\section*{Appendix A. Supplementary Information}
Supplementary information to this article is available.  
%

%
\clearpage

\section*{Appendix A. Supplementary Information}
\setcounter{section}{0}

\newcounter{suppsection}
\renewcommand{\thesection}{S\arabic{section}}

\newcounter{suppfigure}
\renewcommand{\thefigure}{S\arabic{suppfigure}}

\newcounter{supptable}
\renewcommand{\thetable}{S\arabic{supptable}}

\tableofcontents

\addtocontents{toc}{\protect\setcounter{tocdepth}{2}}
\refstepcounter{suppsection}
\section{Structural Properties}
\label{sec:strucprop_si}
\refstepcounter{suppfigure}
\begin{figure*}[h]
    \centering
     \includegraphics[width=0.7\textwidth]{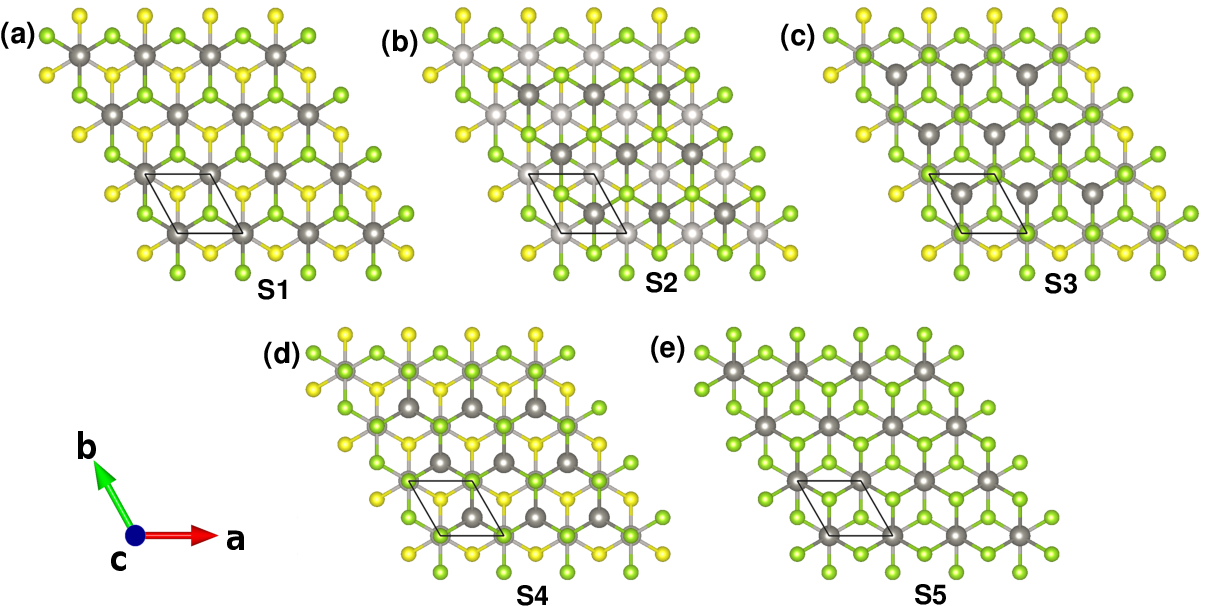}
     \caption{Top view of the considered stacking orders S1-S5 shown using PtWSSe HS in IC1 atom facing type. Black, grey, green and yellow spheres represent W, Pt, S and Se atoms, respectively.}
     \label{fig:sistcking_c}
\end{figure*}
\refstepcounter{suppfigure}
\begin{figure*}[h]
    \centering
     \includegraphics[width=0.7\textwidth]{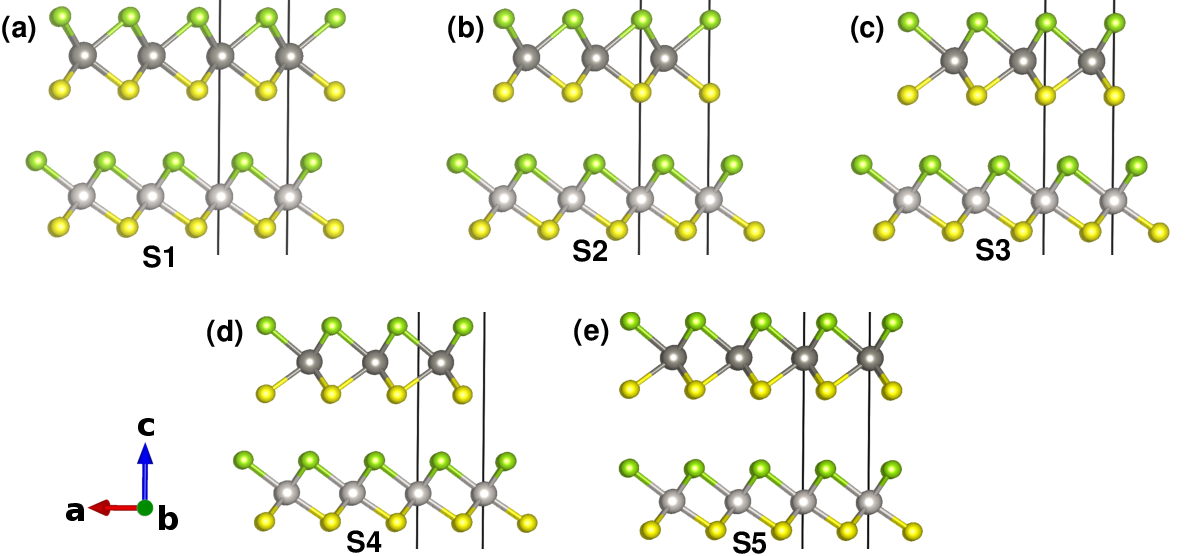}
     \caption{Side view of the considered stacking orders S1-S5 shown using PtWSSe HS in IC1 atom facing type. Black, grey, green and yellow balls represent W, Pt, S and Se atoms, respectively.}
     \label{fig:sistacking_b}
\end{figure*}
Depending on the relative atomic positions of the top and bottom monolayers, different stackings are possible when constructing a bilayer. 
The top and side views of the considered five stacking orders  are shown in \autoref{fig:sistcking_c} and \autoref{fig:sistacking_b}, respectively, using PtWSSe HS in IC1 atom facing type as a representative case. 
Similarly, the same five stacking orders are considered in PtWSTe and PtWSeTe HSs for all four ICs.
In both S1 and S5 stacking orders, alignment of WXY layer (upper layer) and PtSSe layer (lower layer) is such that the metal atoms W and Pt are on top of each other. 
They differ only in the positions of the chalcogen atoms of the WXY layer with respect to the chalcogen atoms of the PtSSe layer.
In stacking order S2, the alignment is such that W (chalcogen) atoms of the WXY layer are on top of upper chalcogen atoms (lower chalcogen atoms) of the PtSSe layer. 
Similarly in S3, W atoms (chalcogen atoms) of the WXY layer are on top of lower chalcogen atoms (Pt atoms) of the PtSSe layer.
The stacking order S4 differs from S2 such that the chalcogen atoms of the WXY layer are on top of Pt atoms of the PtSSe layer.

\refstepcounter{supptable}
\begin{table*}[h!]
  \centering
  \caption{Optimized lattice constants ($a = b$) and interlayer distance $d$ [\AA{}] of three selected HSs in different interface configurations and stacking order S2.}
  \label{tab:lat_ldists2_si}
  \resizebox{\textwidth}{!}{%
  \begin{tabular}{c c c c c c c}
    \toprule
    \multirow{2}{*}{Atom facing types} & 
    \multicolumn{2}{c}{PtWSSe} & 
    \multicolumn{2}{c}{PtWSTe} & 
    \multicolumn{2}{c}{PtWSeTe} \\
    \cmidrule(lr){2-3} \cmidrule(lr){4-5} \cmidrule(lr){6-7}
     & Lattice constant & $d$ & Lattice constant & $d$  & Lattice constant & $d$ \\
    \midrule
    IC1  & 3.4121 & 3.05 (Se-S) & 3.4833 & 2.98 (Se-S) & 3.5244 & 3.00 (Se-Se) \\
    IC2  & 3.4123 & 3.16 (Se-Se) & 3.4832 & 3.30 (Se-Te) & 3.5238 & 3.23 (Se-Te) \\
    IC3  & 3.4123 & 2.95 (S-Se) & 3.4840 & 3.00 (S-Te) & 3.5241 & 2.94 (S-Te) \\
    IC4  & 3.4114 & 2.91 (S-S) & 3.4821 & 2.86 (S-S) & 3.5223 & 2.84  (S-Se) \\
    \bottomrule
  \end{tabular}}
\end{table*}
\refstepcounter{supptable}
\begin{table*}[h!]
  \centering
  \caption{Optimized lattice constants ($a = b$) and interlayer distance $d$ [\AA{}] of three selected HSs in different interface configurations and stacking order S3.}
  \label{tab:lat_ldists3_si}
  \resizebox{\textwidth}{!}{%
  \begin{tabular}{c c c c c c c}
    \toprule
    \multirow{2}{*}{Atom facing types} & 
    \multicolumn{2}{c}{PtWSSe} & 
    \multicolumn{2}{c}{PtWSTe} & 
    \multicolumn{2}{c}{PtWSeTe} \\
    \cmidrule(lr){2-3} \cmidrule(lr){4-5} \cmidrule(lr){6-7}
     & Lattice constant & $d$ & Lattice constant & $d$  & Lattice constant & $d$ \\
    \midrule
    IC1  & 3.4114 & 3.00 (Se-S) & 3.4825 & 2.93 (Se-S) & 3.5228 & 3.01 (Se-Se) \\
    IC2  & 3.4113 & 3.14 (Se-Se) & 3.4819 & 3.32 (Se-Te) & 3.5221 & 3.26 (Se-Te) \\
    IC3  & 3.4114 & 3.01 (S-Se) & 3.4829 & 3.15 (S-Te) & 3.5221 & 3.10 (S-Te) \\
    IC4  & 3.4111 & 2.90 (S-S) & 3.4817 & 2.85 (S-S) & 3.5220 & 2.90  (S-Se) \\
    \bottomrule
  \end{tabular}}
\end{table*}
\refstepcounter{supptable}
\begin{table*}[h!]
  \centering
  \caption{Optimized lattice constants ($a = b$) and interlayer distance $d$ [\AA{}] of three selected HSs in different interface configurations and stacking order S4.}
  \label{tab:lat_ldists4_si}
  \resizebox{\textwidth}{!}{%
  \begin{tabular}{c c c c c c c}
    \toprule
    \multirow{2}{*}{Atom facing types} & 
    \multicolumn{2}{c}{PtWSSe} & 
    \multicolumn{2}{c}{PtWSTe} & 
    \multicolumn{2}{c}{PtWSeTe} \\
    \cmidrule(lr){2-3} \cmidrule(lr){4-5} \cmidrule(lr){6-7}
     & Lattice constant & $d$ & Lattice constant & $d$  & Lattice constant & $d$ \\
    \midrule
    IC1  & 3.4089 & 3.07 (Se-S) & 3.4794 & 3.04 (Se-S) & 3.5194 & 3.09 (Se-Se) \\
    IC2  & 3.4092 & 3.16 (Se-Se) & 3.4795 & 3.28 (Se-Te) & 3.5191 & 3.25 (Se-Te) \\
    IC3  & 3.4086 & 3.05 (S-Se) & 3.4788 & 3.11 (S-Te) & 3.5177 & 3.09 (S-Te) \\
    IC4  & 3.4084 & 3.00 (S-S) & 3.4791 & 2.98 (S-S) & 3.5187 & 3.00  (S-Se) \\
    \bottomrule
  \end{tabular}}
\end{table*}
\refstepcounter{supptable}
\begin{table*}[h!]
  \centering
  \caption{Optimized lattice constants ($a = b$) and interlayer distance $d$ [\AA{}] of three selected HSs in different interface configurations and stacking order S5.}
  \label{tab:lat_ldists5_si}
  \resizebox{\textwidth}{!}{%
  \begin{tabular}{c c c c c c c}
    \toprule
    \multirow{2}{*}{Atom facing types} & 
    \multicolumn{2}{c}{PtWSSe} & 
    \multicolumn{2}{c}{PtWSTe} & 
    \multicolumn{2}{c}{PtWSeTe} \\
    \cmidrule(lr){2-3} \cmidrule(lr){4-5} \cmidrule(lr){6-7}
     & Lattice constant & $d$ & Lattice constant & $d$  & Lattice constant & $d$ \\
    \midrule
    IC1  & 3.4135 & 3.06 (Se-S) & 3.4864 & 2.93 (Se-S) & 3.5268 & 3.01 (Se-Se) \\
    IC2  & 3.4125 & 3.23 (Se-Se) & 3.4825 & 3.44 (Se-Te) & 3.5234 & 3.38 (Se-Te) \\
    IC3  & 3.4144 & 3.01 (S-Se) & 3.4852 & 3.18 (S-Te) & 3.5264 & 3.07 (S-Te) \\
    IC4  & 3.4147 & 2.86 (S-S) & 3.4868 & 2.73 (S-S) & 3.5282 & 2.76  (S-Se) \\
    \bottomrule
  \end{tabular}}
\end{table*}
Optimized lattice constants and interlayer distances $d$ for stacking orders S2-S5 for all the ICs are reported in Tables \autoref{tab:lat_ldists2_si}-\autoref{tab:lat_ldists5_si}.
Here, $d$ is the distance between the chalcogen atoms at the interface of a HS.
The tables show that varying the interfacial atom types has little influence on the lattice constants but has greater effects on the interlayer distance due to differences in atomic size.
\section{Stability}
\label{sec:stability_si}
\refstepcounter{supptable}
\begin{table*}[ht]
  \centering
  \caption{The layer binding energy $E_b$ in $\mathrm{eV}$ for PtWSTe HSs with different ICs and stacking orders.}
  \label{tab:Eb_si}
  \begin{tabular}{c c c c c }
    \toprule
    \multirow{2}{*}{Stacking order} & 
    \multicolumn{4}{c}{PtWSTe}  \\
    \cmidrule(lr){2-5}
      & IC1 (Se-S) & IC2 (Se-Te) & IC3 (S-Te) & IC4 (S-S) \\ 
    \midrule
    S1  & 0.060 & 0.059  & 0.053 & 0.077  \\ 
    S2  & -0.031 & -0.026  & -0.056 & -0.023  \\
    S3  & -0.037 & -0.025  & -0.040 & -0.020  \\
    S4  & -0.024 & -0.034  & -0.051 & -0.005  \\
    S5  & -0.028 & -0.003  & -0.027 & -0.024  \\ 
    \bottomrule
  \end{tabular}
\end{table*}
\refstepcounter{suppfigure}
\begin{figure*}[h]
    \centering
     \includegraphics[width=0.9\textwidth]{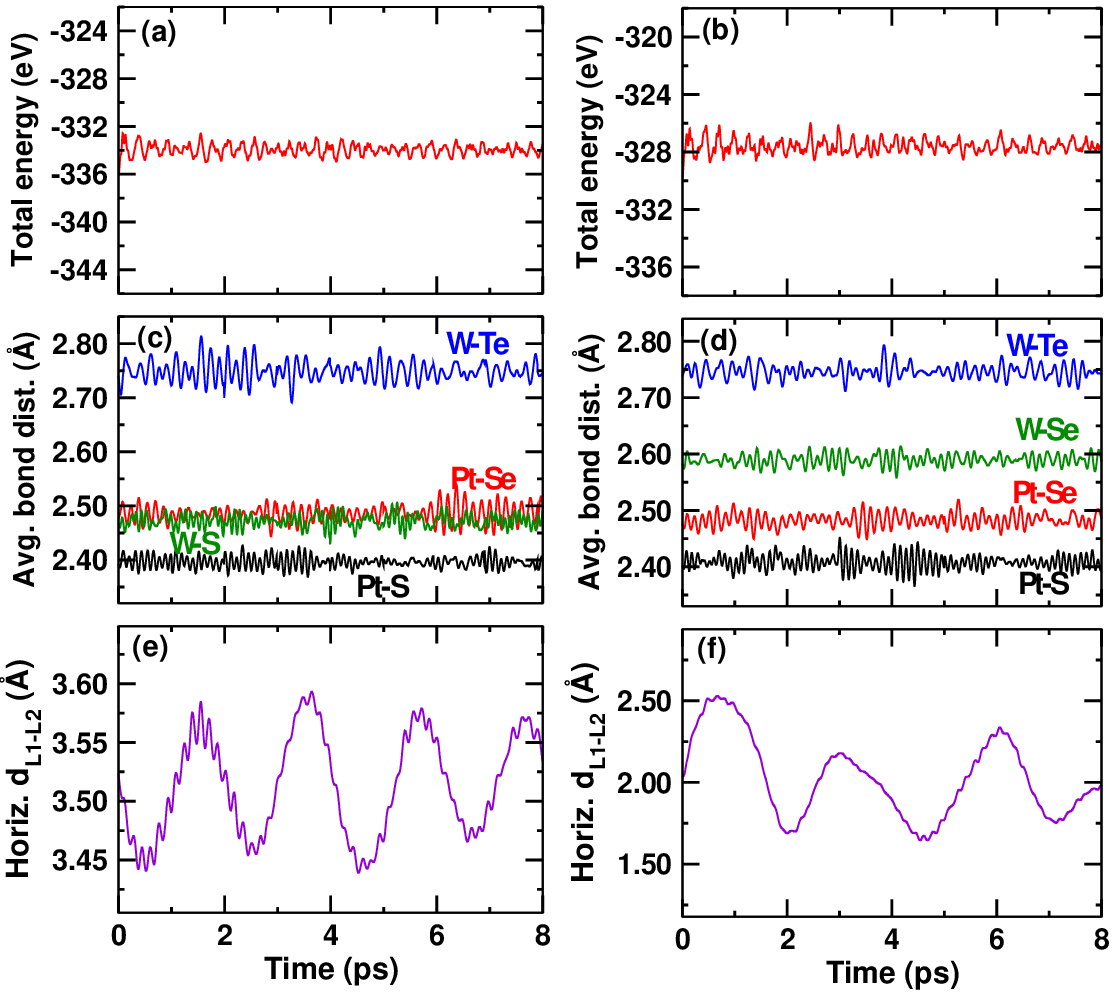}
     \caption{Variation of total energy of the system, average bond distance (Avg. bond dist.) and horizontal distance between layers (Horizon. d$_{\mathrm{L1-L2}}$) as a function of time. Left and right panels show the results for PtWSTe-IC1-S3 and PtWSeTe-IC2-S2 systems, respectively.}
     \label{fig:213_322aimd_si}
\end{figure*}
The layer binding energy $E_{b}$ for PtWSTe in different atom facing types (IC) and stacking orders are presented in \autoref{tab:Eb_si}.
$E_{b}$ values are negative for all stacking orders except S1, indicating that the formation of the bilayer is energetically favourable.
The total energy variation, average bond distance and horizontal distance between layers as a function of simulation time are shown for PtWSTe-IC1-S3 and PtWSeTe-IC2-S2 HSs in \autoref{fig:213_322aimd_si}. 
The variation in the average bond distance between the metal and chalcogen atoms is less than 0.1 \AA{}.
The change in the horizontal distance between the layers (max $\sim$0.5 \AA{}) suggests no significant relative layer sliding with respect the initial stacking order.
Thus, results reported in \autoref{fig:213_322aimd_si} indicate that PtWSTe-IC1-S3 and PtWSeTe-IC2-S2 systems are thermally stable.
\section{Electronic properties}
\label{sec:elecprop_si}
\refstepcounter{suppfigure}
\begin{figure}[ht]
    \centering
     \includegraphics[width=0.4\textwidth]{213_si}
     \caption{Electronic band structure of  PtWSTe HS in atom facing type IC1 and stacking order S3 calculated using the HSE06 functional.}
     \label{fig:213si}
\end{figure}
\refstepcounter{suppfigure}
\begin{figure}[ht]
    \centering
     \includegraphics[width=0.4\textwidth]{322_si}
     \caption{Electronic band structure of  PtWSeTe HS in atom facing type IC2 and stacking order S2 calculated using the HSE06 functional.}
     \label{fig:322si}
\end{figure}
\autoref{fig:213si} shows the electronic band structure of PtWSTe HS in IC1-S3 configuration obtained from the HSE06 hybrid functional calculations. 
It is an indirect band gap semiconductor with gap value of 0.57 $\mathrm{eV}$.
The VBM is at the $\Gamma$-point and the CBM is realised at the $K$-point.
The electronic band structure of PtWSeTe in IC2-S2 configuration without external strain (\autoref{fig:322si}) shows that this HS has indirect band gap with width equal to 1.22 $\mathrm{eV}$.
PtWSeTe in IC2-S2 configuration has other conduction band minima with close energy values along $\Gamma$-$K$ and $\Gamma$-$M$ directions, suggesting enhanced efficiency for electron photoexcitation.

\section{Band edge tuning by strain}
\label{sec:strain_si}
\refstepcounter{suppfigure}
\begin{figure*}[ht]
    \centering
     \includegraphics[width=0.9 \textwidth]{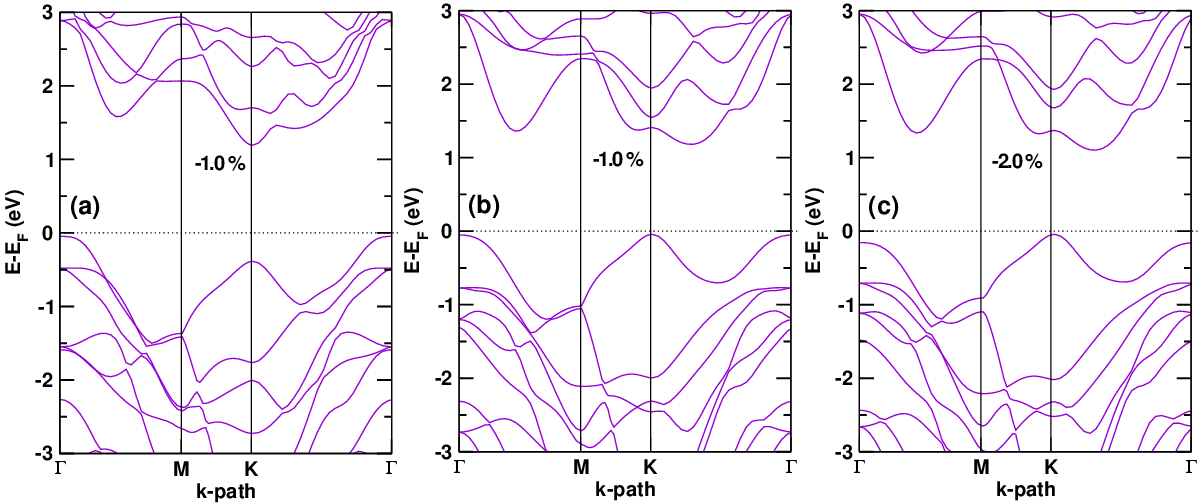}
     \caption{Electronic band structures at different biaxial compressive strains: (a) PtWSeTe-IC1-S2, -1.0 $\%$; (b) PtWSeTe-IC2-S2, -1.0 $\%$; (c) PtWSeTe-IC2-S2, -2.0 $\%$.}
     \label{fig:eband_strain_si}
\end{figure*}
\refstepcounter{suppfigure}
\begin{figure*}[ht]
    \centering
     \includegraphics[width=0.9\textwidth]{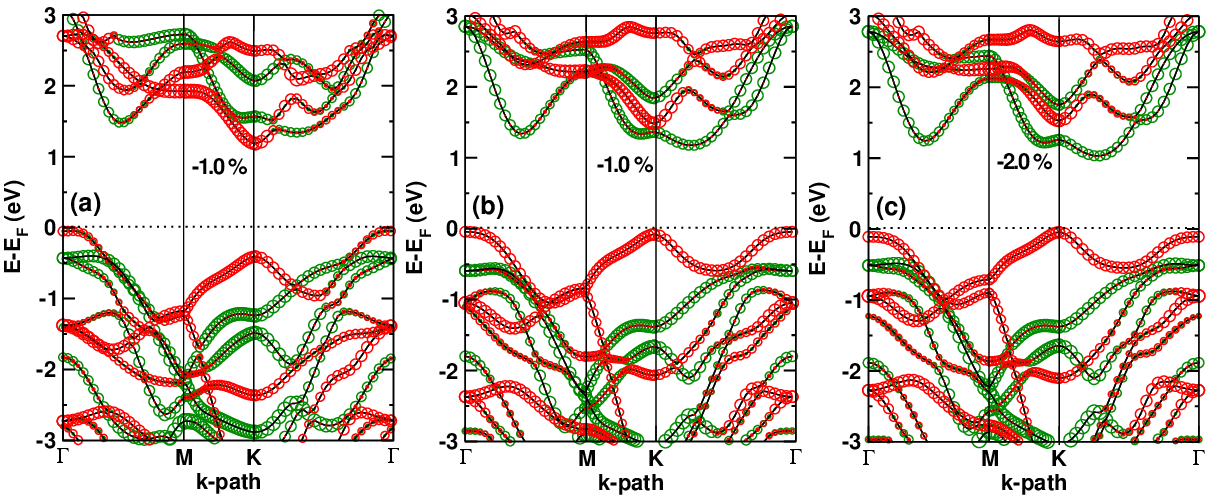}
     \caption{Layer projected band structures at different strain values of (a) PtWSeTe-IC1-S2 (-1.0 $\%$), (b) PtWSeTe-IC2-S2 (-1.0 $\%$) and (c) PtWSeTe-IC1-S2 (-2.0 $\%$). Red and green circles represent projection onto WSeTe and PtSSe component monolayers, respectively, while larger circles indicate larger monolayer contributions.}
     \label{fig:lp_strain_si}
\end{figure*}
\refstepcounter{suppfigure}
\begin{figure*}[ht]
    \centering
     \includegraphics[width=17cm,height=7cm]{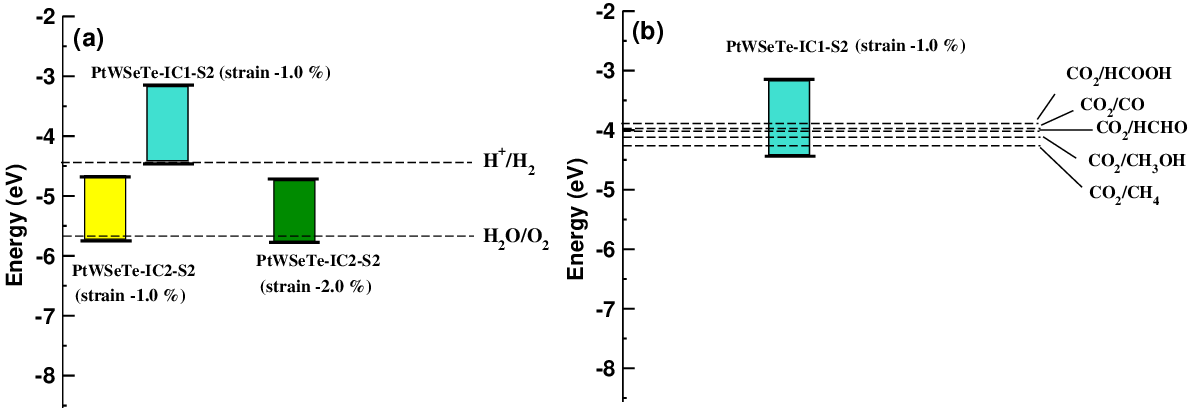}
     \caption{Alignment of band edge positions of strained heterostructures with (a) water redox potential and (b) CO$_2$ reduction potential.}
     \label{fig:redox_si}
\end{figure*}
The shift in the energy positions of the electronic bands edges for PtWSeTe-IC1-S2 (-1.0 $\% $ strain) and  PtWSeTe-IC2-S2 (-1.0 and -2.0 $\%$ strain) by the application of compressive biaxial strain can be observed from the electronic band structures in \autoref{fig:eband_strain_si}.
The corresponding component layer characters are shown as layer projected band structures in \autoref{fig:lp_strain_si}.  
These plots suggest that strained PtWSeTe-IC1-S2 has type-I band alignment and PtWSeTe-IC2-S2 has type-II band alignment.
The photocatalytic activity of the strained HSs is assessed by aligning the band edges with respect to the redox potentials;
the relative energy levels are shown in \autoref{fig:redox_si}.
By inspecting the figure, it can be observed that the valence band edges of strained PtWSeTe-IC2-S2 are suitably positioned to accept electrons for the oxidation of water molecules, indicating its potential as a candidate for the OER;
instead, the conduction band edge of strained PtWSeTe-IC1-S2 lies at a favourable level to donate electrons for the HER and CO$_{2}$ reduction reactions.
\clearpage

\section{Optimized structure files}
\label{sec:poscar_si}
Here, we provide the geometry optimized structure files of the component monolayers and HSs of  PtWSSe, PtWSTe and PtWSeTe HSs in four atom facing types (ICs) and five stacking orders considered in \textsc{POSCAR} format.
s%
\subsection{1T-PtSSe monolayer}
\label{subsec:ptsse_ml}
\begin{Verbatim}
1T-PtSSe                                
   1.00000000000000     
     3.6324779499665327    0.0000000000000000    0.0000000000000000
    -1.8162389749832664    3.1458181836204857    0.0000000000000000
     0.0000000000000000    0.0000000000000000   30.0000000000000000
   S    Se   Pt
     1     1     1
Direct
  0.3333333429999996  0.6666666870000029  0.4578453771697042
  0.6666666870000029  0.3333333429999996  0.5436687222117342
  0.0000000000000000  0.0000000000000000  0.4984859306185783
 
\end{Verbatim}
\subsection{2H-WSSe monolayer}
\label{subsec:wsse_ml}
\begin{Verbatim}
2H-WSSe                                 
   1.00000000000000     
     3.2607638454095018    0.0000000000000000    0.0000000000000000
    -1.6303819227047509    2.8239043255661351    0.0000000000000000
     0.0000000000000000    0.0000000000000000   30.0000000000000000
   W    Se   S 
     1     1     1
Direct
  0.6666666870000029  0.3333333429999996  0.4980307788903815
  0.3333333429999996  0.6666666870000029  0.5549485235646472
  0.3333333429999996  0.6666666870000029  0.4470206975449855

\end{Verbatim}
\subsection{2H-WSTe monolayer}
\label{subsec:wste_ml}
\begin{Verbatim}
2H-WSTe                                 
   1.00000000000000     
     3.3796362794363404    0.0000000000000000    0.0000000000000000
    -1.6898181397181702    2.9268508738692156    0.0000000000000000
     0.0000000000000000    0.0000000000000000   30.0000000000000000
   W    Te   S 
     1     1     1
Direct
  0.6666666870000029  0.3333333429999996  0.4952511731597298
  0.3333333429999996  0.6666666870000029  0.5585934497671659
  0.3333333429999996  0.6666666870000029  0.4461553770731186

\end{Verbatim}
\subsection{2H-WSeTe monolayer}
\label{subsec:wsete_ml}
\begin{Verbatim}
2H-WSeTe                                
   1.00000000000000     
     3.4539589685627483    0.0000000000000000    0.0000000000000000
    -1.7269794842813742    2.9912162104206517    0.0000000000000000
     0.0000000000000000    0.0000000000000000   30.0000000000000000
   W    Te   Se
     1     1     1
Direct
  0.6666666870000029  0.3333333429999996  0.4972823734109824
  0.3333333429999996  0.6666666870000029  0.5593110971777620
  0.3333333429999996  0.6666666870000029  0.4434065294112557
 
  0.00000000E+00  0.00000000E+00  0.00000000E+00
  0.00000000E+00  0.00000000E+00  0.00000000E+00
  0.00000000E+00  0.00000000E+00  0.00000000E+00
\end{Verbatim}
\subsection{PtWSSe-IC1-S1}
\label{susec:111}
\begin{Verbatim}
1T-PtSSe/2H-WSSe IC1-S1                       
   1.00000000000000     
     3.4062991490817032    0.0000000000000000    0.0000000000000000
    -1.7031495745408516    2.9499415961096962    0.0000000000000000
     0.0000000000000000    0.0000000000000000   30.0000000000000000
   S    Se   Pt   W    Se   S 
     1     1     1     1     1     1
Direct
  0.3333333429999996  0.6666666870000029  0.4400113977459128
  0.6666666870000029  0.3333333429999996  0.5348210427445892
  0.0000000000000000  0.0000000000000000  0.4844788826627706
  0.0000000000000000  0.0000000000000000  0.7048641662874715
  0.6666666870000029  0.3333333429999996  0.7602064451696151
  0.6666666870000029  0.3333333429999996  0.6555180963896277

\end{Verbatim}
\subsection{PtWSSe-IC1-S2}
\label{susec:112}
\begin{Verbatim}
1T-PtSSe/2H-WSSe IC1-S2                       
   1.00000000000000     
     3.4121086780438374    0.0000000000000000    0.0000000000000000
    -1.7060543390219187    2.9549727957756775    0.0000000000000000
     0.0000000000000000    0.0000000000000000   30.0000000000000000
   S    Se   Pt   W    Se   S 
     1     1     1     1     1     1
Direct
  0.3333333429999996  0.6666666870000029  0.4498218859112200
  0.6666666870000029  0.3333333429999996  0.5441735178287743
  0.0000000000000000  0.0000000000000000  0.4943048420909193
  0.6666666870000029  0.3333333429999996  0.6951972024737358
  0.3333333429999996  0.6666666870000029  0.7504609811066985
  0.3333333429999996  0.6666666870000029  0.6459416015886603
 
  0.00000000E+00  0.00000000E+00  0.00000000E+00
  0.00000000E+00  0.00000000E+00  0.00000000E+00
  0.00000000E+00  0.00000000E+00  0.00000000E+00
  0.00000000E+00  0.00000000E+00  0.00000000E+00
  0.00000000E+00  0.00000000E+00  0.00000000E+00
  0.00000000E+00  0.00000000E+00  0.00000000E+00
\end{Verbatim}
\subsection{PtWSSe-IC1-S3}
\label{susec:113}
\begin{Verbatim}
1T-PtSSe/2H-WSSe IC1-S3                       
   1.00000000000000     
     3.4114423434093850    0.0000000000000000    0.0000000000000000
    -1.7057211717046925    2.9543957330537856    0.0000000000000000
     0.0000000000000000    0.0000000000000000   30.0000000000000000
   S    Se   Pt   W    Se   S 
     1     1     1     1     1     1
Direct
  0.3333333429999996  0.6666666870000029  0.4506352076647246
  0.6666666870000029  0.3333333429999996  0.5450589338692637
  0.0000000000000000  0.0000000000000000  0.4950681098358061
  0.3333333429999996  0.6666666870000029  0.6943683643531884
  0.0000000000000000  0.0000000000000000  0.7496445116831865
  0.0000000000000000  0.0000000000000000  0.6451249035938389
 
  0.00000000E+00  0.00000000E+00  0.00000000E+00
  0.00000000E+00  0.00000000E+00  0.00000000E+00
  0.00000000E+00  0.00000000E+00  0.00000000E+00
  0.00000000E+00  0.00000000E+00  0.00000000E+00
  0.00000000E+00  0.00000000E+00  0.00000000E+00
  0.00000000E+00  0.00000000E+00  0.00000000E+00
\end{Verbatim}
\subsection{PtWSSe-IC1-S4}
\label{susec:114}
\begin{Verbatim}
1T-PtSSe/2H-WSSe IC1-S4                       
   1.00000000000000     
     3.4089148106985672    0.0000000000000000    0.0000000000000000
    -1.7044574053492836    2.9522068255179410    0.0000000000000000
     0.0000000000000000    0.0000000000000000   30.0000000000000000
   S    Se   Pt   W    Se   S 
     1     1     1     1     1     1
Direct
  0.3333333429999996  0.6666666870000029  0.4492951204522342
  0.6666666870000029  0.3333333429999996  0.5438577039608816
  0.0000000000000000  0.0000000000000000  0.4937802769217114
  0.6666666870000029  0.3333333429999996  0.6956588856599595
  0.0000000000000000  0.0000000000000000  0.7509542992434106
  0.0000000000000000  0.0000000000000000  0.6463537447618251
 
  0.00000000E+00  0.00000000E+00  0.00000000E+00
  0.00000000E+00  0.00000000E+00  0.00000000E+00
  0.00000000E+00  0.00000000E+00  0.00000000E+00
  0.00000000E+00  0.00000000E+00  0.00000000E+00
  0.00000000E+00  0.00000000E+00  0.00000000E+00
  0.00000000E+00  0.00000000E+00  0.00000000E+00
\end{Verbatim}
\subsection{PtWSSe-IC1-S5}
\label{susec:115}
\begin{Verbatim}
1T-PtSSe/2H-WSSe IC1-S5                       
   1.00000000000000     
     3.4135179808941514    0.0000000000000000    0.0000000000000000
    -1.7067589904470757    2.9561932878458452    0.0000000000000000
     0.0000000000000000    0.0000000000000000   30.0000000000000000
   S    Se   Pt   W    Se   S 
     1     1     1     1     1     1
Direct
  0.3333333429999996  0.6666666870000029  0.4496720044234337
  0.6666666870000029  0.3333333429999996  0.5439667045848822
  0.0000000000000000  0.0000000000000000  0.4941408840746107
  0.0000000000000000  0.0000000000000000  0.6953544828239515
  0.3333333429999996  0.6666666870000029  0.7506221024915050
  0.3333333429999996  0.6666666870000029  0.6461438526016252
 
  0.00000000E+00  0.00000000E+00  0.00000000E+00
  0.00000000E+00  0.00000000E+00  0.00000000E+00
  0.00000000E+00  0.00000000E+00  0.00000000E+00
  0.00000000E+00  0.00000000E+00  0.00000000E+00
  0.00000000E+00  0.00000000E+00  0.00000000E+00
  0.00000000E+00  0.00000000E+00  0.00000000E+00
\end{Verbatim}
\subsection{PtWSSe-IC2-S1}
\label{susec:121}
\begin{Verbatim}
1T-PtSSe/2H-WSSe IC2-S1                        
   1.00000000000000     
     3.4060663780568809    0.0000000000000000    0.0000000000000000
    -1.7030331890284405    2.9497400104892733    0.0000000000000000
     0.0000000000000000    0.0000000000000000   30.0000000000000000
   S    Se   Pt   W    S    Se
     1     1     1     1     1     1
Direct
  0.3333333429999996  0.6666666870000029  0.4373426918871459
  0.6666666870000029  0.3333333429999996  0.5321621866941157
  0.0000000000000000  0.0000000000000000  0.4818088676420587
  0.0000000000000000  0.0000000000000000  0.7114648601037601
  0.6666666870000029  0.3333333429999996  0.7609279139728358
  0.6666666870000029  0.3333333429999996  0.6561935107000778
 
  0.00000000E+00  0.00000000E+00  0.00000000E+00
  0.00000000E+00  0.00000000E+00  0.00000000E+00
  0.00000000E+00  0.00000000E+00  0.00000000E+00
  0.00000000E+00  0.00000000E+00  0.00000000E+00
  0.00000000E+00  0.00000000E+00  0.00000000E+00
  0.00000000E+00  0.00000000E+00  0.00000000E+00
\end{Verbatim}
\subsection{PtWSSe-IC2-S2}
\label{susec:122}
\begin{Verbatim}
1T-PtSSe/2H-WSSe IC2-S2                        
   1.00000000000000     
     3.4123609106120454    0.0000000000000000    0.0000000000000000
    -1.7061804553060227    2.9551912355875856    0.0000000000000000
     0.0000000000000000    0.0000000000000000   30.0000000000000000
   S    Se   Pt   W    S    Se
     1     1     1     1     1     1
Direct
  0.3333333429999996  0.6666666870000029  0.4470758757069362
  0.6666666870000029  0.3333333429999996  0.5414194987356993
  0.0000000000000000  0.0000000000000000  0.4915570205331932
  0.6666666870000029  0.3333333429999996  0.7018748910849055
  0.3333333429999996  0.6666666870000029  0.7512653942595264
  0.3333333429999996  0.6666666870000029  0.6467073506797121
 
  0.00000000E+00  0.00000000E+00  0.00000000E+00
  0.00000000E+00  0.00000000E+00  0.00000000E+00
  0.00000000E+00  0.00000000E+00  0.00000000E+00
  0.00000000E+00  0.00000000E+00  0.00000000E+00
  0.00000000E+00  0.00000000E+00  0.00000000E+00
  0.00000000E+00  0.00000000E+00  0.00000000E+00
\end{Verbatim}
\subsection{PtWSSe-IC2-S3}
\label{susec:123}
\begin{Verbatim}
1T-PtSSe/2H-WSSe IC2-S3                       
   1.00000000000000     
     3.4113507138431829    0.0000000000000000    0.0000000000000000
    -1.7056753569215914    2.9543163795176013    0.0000000000000000
     0.0000000000000000    0.0000000000000000   30.0000000000000000
   S    Se   Pt   W    S    Se
     1     1     1     1     1     1
Direct
  0.3333333429999996  0.6666666870000029  0.4473607940996871
  0.6666666870000029  0.3333333429999996  0.5418166840885448
  0.0000000000000000  0.0000000000000000  0.4917771775261954
  0.3333333429999996  0.6666666870000029  0.7015620525490789
  0.0000000000000000  0.0000000000000000  0.7509721214473402
  0.0000000000000000  0.0000000000000000  0.6464112012891405
 
  0.00000000E+00  0.00000000E+00  0.00000000E+00
  0.00000000E+00  0.00000000E+00  0.00000000E+00
  0.00000000E+00  0.00000000E+00  0.00000000E+00
  0.00000000E+00  0.00000000E+00  0.00000000E+00
  0.00000000E+00  0.00000000E+00  0.00000000E+00
  0.00000000E+00  0.00000000E+00  0.00000000E+00
\end{Verbatim}
\subsection{PtWSSe-IC2-S4}
\label{susec:124}
\begin{Verbatim}
1T-PtSSe/2H-WSSe IC2-S4                        
   1.00000000000000     
     3.4092153835998689    0.0000000000000000    0.0000000000000000
    -1.7046076917999344    2.9524671292857425    0.0000000000000000
     0.0000000000000000    0.0000000000000000   30.0000000000000000
   S    Se   Pt   W    S    Se
     1     1     1     1     1     1
Direct
  0.3333333429999996  0.6666666870000029  0.4468192918193239
  0.6666666870000029  0.3333333429999996  0.5413634643792733
  0.0000000000000000  0.0000000000000000  0.4912985656244544
  0.6666666870000029  0.3333333429999996  0.7020702204617422
  0.0000000000000000  0.0000000000000000  0.7514954561689535
  0.0000000000000000  0.0000000000000000  0.6468530325462893
 
  0.00000000E+00  0.00000000E+00  0.00000000E+00
  0.00000000E+00  0.00000000E+00  0.00000000E+00
  0.00000000E+00  0.00000000E+00  0.00000000E+00
  0.00000000E+00  0.00000000E+00  0.00000000E+00
  0.00000000E+00  0.00000000E+00  0.00000000E+00
  0.00000000E+00  0.00000000E+00  0.00000000E+00
\end{Verbatim}
\subsection{PtWSSe-IC2-S5}
\label{susec:125}
\begin{Verbatim}
1T-PtSSe/2H-WSSe IC2-S5                        
   1.00000000000000     
     3.4125574287041780    0.0000000000000000    0.0000000000000000
    -1.7062787143520890    2.9553614252472760    0.0000000000000000
     0.0000000000000000    0.0000000000000000   30.0000000000000000
   S    Se   Pt   W    S    Se
     1     1     1     1     1     1
Direct
  0.3333333429999996  0.6666666870000029  0.4458386223338024
  0.6666666870000029  0.3333333429999996  0.5402169722193193
  0.0000000000000000  0.0000000000000000  0.4903050828857900
  0.0000000000000000  0.0000000000000000  0.7030911769843939
  0.3333333429999996  0.6666666870000029  0.7524954105425863
  0.3333333429999996  0.6666666870000029  0.6479527660340949
 
  0.00000000E+00  0.00000000E+00  0.00000000E+00
  0.00000000E+00  0.00000000E+00  0.00000000E+00
  0.00000000E+00  0.00000000E+00  0.00000000E+00
  0.00000000E+00  0.00000000E+00  0.00000000E+00
  0.00000000E+00  0.00000000E+00  0.00000000E+00
  0.00000000E+00  0.00000000E+00  0.00000000E+00
\end{Verbatim}
\subsection{PtWSSe-IC3-S1}
\label{susec:131}
\begin{Verbatim}
1T-PtSSe/2H-WSSe IC3-S1                        
   1.00000000000000     
     3.4069369764789332    0.0000000000000000    0.0000000000000000
    -1.7034684882394666    2.9504939707793811    0.0000000000000000
     0.0000000000000000    0.0000000000000000   30.0000000000000000
   Se   S    Pt   W    S    Se
     1     1     1     1     1     1
Direct
  0.3333333429999996  0.6666666870000029  0.4375431469071955
  0.6666666870000029  0.3333333429999996  0.5322602173693980
  0.0000000000000000  0.0000000000000000  0.4879445677031669
  0.0000000000000000  0.0000000000000000  0.7093311151160719
  0.6666666870000029  0.3333333429999996  0.7587784461433742
  0.6666666870000029  0.3333333429999996  0.6540425377608088
 
  0.00000000E+00  0.00000000E+00  0.00000000E+00
  0.00000000E+00  0.00000000E+00  0.00000000E+00
  0.00000000E+00  0.00000000E+00  0.00000000E+00
  0.00000000E+00  0.00000000E+00  0.00000000E+00
  0.00000000E+00  0.00000000E+00  0.00000000E+00
  0.00000000E+00  0.00000000E+00  0.00000000E+00
\end{Verbatim}
\subsection{PtWSSe-IC3-S2}
\label{susec:132}
\begin{Verbatim}
1T-PtSSe/2H-WSSe IC3-S2                       
   1.00000000000000     
     3.4123619301387040    0.0000000000000000    0.0000000000000000
    -1.7061809650693520    2.9551921185232488    0.0000000000000000
     0.0000000000000000    0.0000000000000000   30.0000000000000000
   Se   S    Pt   W    S    Se
     1     1     1     1     1     1
Direct
  0.3333333429999996  0.6666666870000029  0.4495592071080097
  0.6666666870000029  0.3333333429999996  0.5438698403081617
  0.0000000000000000  0.0000000000000000  0.4999652868181741
  0.6666666870000029  0.3333333429999996  0.6974330807257161
  0.3333333429999996  0.6666666870000029  0.7468105691410472
  0.3333333429999996  0.6666666870000029  0.6422620468989209
 
  0.00000000E+00  0.00000000E+00  0.00000000E+00
  0.00000000E+00  0.00000000E+00  0.00000000E+00
  0.00000000E+00  0.00000000E+00  0.00000000E+00
  0.00000000E+00  0.00000000E+00  0.00000000E+00
  0.00000000E+00  0.00000000E+00  0.00000000E+00
  0.00000000E+00  0.00000000E+00  0.00000000E+00
\end{Verbatim}
\subsection{PtWSSe-IC3-S3}
\label{susec:133}
\begin{Verbatim}
1T-PtSSe/2H-WSSe IC3-S3                        
   1.00000000000000     
     3.4114503080753242    0.0000000000000000    0.0000000000000000
    -1.7057251540376621    2.9544026306569395    0.0000000000000000
     0.0000000000000000    0.0000000000000000   30.0000000000000000
   Se   S    Pt   W    S    Se
     1     1     1     1     1     1
Direct
  0.3333333429999996  0.6666666870000029  0.4485471415557143
  0.6666666870000029  0.3333333429999996  0.5429736143358852
  0.0000000000000000  0.0000000000000000  0.4988992559007173
  0.3333333429999996  0.6666666870000029  0.6984100043697055
  0.0000000000000000  0.0000000000000000  0.7478170006558571
  0.0000000000000000  0.0000000000000000  0.6432530141821289
 
  0.00000000E+00  0.00000000E+00  0.00000000E+00
  0.00000000E+00  0.00000000E+00  0.00000000E+00
  0.00000000E+00  0.00000000E+00  0.00000000E+00
  0.00000000E+00  0.00000000E+00  0.00000000E+00
  0.00000000E+00  0.00000000E+00  0.00000000E+00
  0.00000000E+00  0.00000000E+00  0.00000000E+00
\end{Verbatim}
\subsection{PtWSSe-IC3-S4}
\label{susec:134}
\begin{Verbatim}
1T-PtSSe/2H-WSSe IC3-S4                        
   1.00000000000000     
     3.4086043055874606    0.0000000000000000    0.0000000000000000
    -1.7043021527937303    2.9519379202033402    0.0000000000000000
     0.0000000000000000    0.0000000000000000   30.0000000000000000
   Se   S    Pt   W    S    Se
     1     1     1     1     1     1
Direct
  0.3333333429999996  0.6666666870000029  0.4477591746389820
  0.6666666870000029  0.3333333429999996  0.5423091166265408
  0.0000000000000000  0.0000000000000000  0.4981817734192902
  0.6666666870000029  0.3333333429999996  0.6991576146253351
  0.0000000000000000  0.0000000000000000  0.7485823450063123
  0.0000000000000000  0.0000000000000000  0.6439100066835195
 
  0.00000000E+00  0.00000000E+00  0.00000000E+00
  0.00000000E+00  0.00000000E+00  0.00000000E+00
  0.00000000E+00  0.00000000E+00  0.00000000E+00
  0.00000000E+00  0.00000000E+00  0.00000000E+00
  0.00000000E+00  0.00000000E+00  0.00000000E+00
  0.00000000E+00  0.00000000E+00  0.00000000E+00
\end{Verbatim}
\subsection{PtWSSe-IC3-S5}
\label{susec:135}
\begin{Verbatim}
1T-PtSSe/2H-WSSe IC3-S5                       
   1.00000000000000     
     3.4144084175253666    0.0000000000000000    0.0000000000000000
    -1.7072042087626833    2.9569644285885777    0.0000000000000000
     0.0000000000000000    0.0000000000000000   30.0000000000000000
   Se   S    Pt   W    S    Se
     1     1     1     1     1     1
Direct
  0.3333333429999996  0.6666666870000029  0.4486850313503297
  0.6666666870000029  0.3333333429999996  0.5429062751989093
  0.0000000000000000  0.0000000000000000  0.4990390900339392
  0.0000000000000000  0.0000000000000000  0.6983324879158417
  0.3333333429999996  0.6666666870000029  0.7477102445568988
  0.3333333429999996  0.6666666870000029  0.6432269019440753
 
  0.00000000E+00  0.00000000E+00  0.00000000E+00
  0.00000000E+00  0.00000000E+00  0.00000000E+00
  0.00000000E+00  0.00000000E+00  0.00000000E+00
  0.00000000E+00  0.00000000E+00  0.00000000E+00
  0.00000000E+00  0.00000000E+00  0.00000000E+00
  0.00000000E+00  0.00000000E+00  0.00000000E+00
\end{Verbatim}
\subsection{PtWSSe-IC4-S1}
\label{susec:141}
\begin{Verbatim}
1T-PtSSe/2H-WSSe IC4-S1                       
   1.00000000000000     
     3.4070898385173534    0.0000000000000000    0.0000000000000000
    -1.7035449192586767    2.9506263532477264    0.0000000000000000
     0.0000000000000000    0.0000000000000000   30.0000000000000000
   Se   S    Pt   W    Se   S 
     1     1     1     1     1     1
Direct
  0.3333333429999996  0.6666666870000029  0.4398426742636801
  0.6666666870000029  0.3333333429999996  0.5345564434782162
  0.0000000000000000  0.0000000000000000  0.4902463645107105
  0.0000000000000000  0.0000000000000000  0.7030977539891836
  0.6666666870000029  0.3333333429999996  0.7584226414083162
  0.6666666870000029  0.3333333429999996  0.6537341533498875
 
  0.00000000E+00  0.00000000E+00  0.00000000E+00
  0.00000000E+00  0.00000000E+00  0.00000000E+00
  0.00000000E+00  0.00000000E+00  0.00000000E+00
  0.00000000E+00  0.00000000E+00  0.00000000E+00
  0.00000000E+00  0.00000000E+00  0.00000000E+00
  0.00000000E+00  0.00000000E+00  0.00000000E+00
\end{Verbatim}
\subsection{PtWSSe-IC4-S2}
\label{susec:142}
\begin{Verbatim}
1T-PtSSe/2H-WSSe IC4-S2                       
   1.00000000000000     
     3.4114577311239014    0.0000000000000000    0.0000000000000000
    -1.7057288655619507    2.9544090592063457    0.0000000000000000
     0.0000000000000000    0.0000000000000000   30.0000000000000000
   Se   S    Pt   W    Se   S 
     1     1     1     1     1     1
Direct
  0.3333333429999996  0.6666666870000029  0.4512432935264243
  0.6666666870000029  0.3333333429999996  0.5455925479421779
  0.0000000000000000  0.0000000000000000  0.5016540461956041
  0.6666666870000029  0.3333333429999996  0.6918121583304497
  0.3333333429999996  0.6666666870000029  0.7470652875597992
  0.3333333429999996  0.6666666870000029  0.6425326974455245
 
  0.00000000E+00  0.00000000E+00  0.00000000E+00
  0.00000000E+00  0.00000000E+00  0.00000000E+00
  0.00000000E+00  0.00000000E+00  0.00000000E+00
  0.00000000E+00  0.00000000E+00  0.00000000E+00
  0.00000000E+00  0.00000000E+00  0.00000000E+00
  0.00000000E+00  0.00000000E+00  0.00000000E+00
\end{Verbatim}
\subsection{PtWSSe-IC4-S3}
\label{susec:143}
\begin{Verbatim}
1T-PtSSe/2H-WSSe IC4-S3                       
   1.00000000000000     
     3.4110814468859192    0.0000000000000000    0.0000000000000000
    -1.7055407234429596    2.9540831874964382    0.0000000000000000
     0.0000000000000000    0.0000000000000000   30.0000000000000000
   Se   S    Pt   W    Se   S 
     1     1     1     1     1     1
Direct
  0.3333333429999996  0.6666666870000029  0.4513215389491307
  0.6666666870000029  0.3333333429999996  0.5457371401465139
  0.0000000000000000  0.0000000000000000  0.5016880808291404
  0.3333333429999996  0.6666666870000029  0.6917092076065074
  0.0000000000000000  0.0000000000000000  0.7469842099787485
  0.0000000000000000  0.0000000000000000  0.6424598534899673
 
  0.00000000E+00  0.00000000E+00  0.00000000E+00
  0.00000000E+00  0.00000000E+00  0.00000000E+00
  0.00000000E+00  0.00000000E+00  0.00000000E+00
  0.00000000E+00  0.00000000E+00  0.00000000E+00
  0.00000000E+00  0.00000000E+00  0.00000000E+00
  0.00000000E+00  0.00000000E+00  0.00000000E+00
\end{Verbatim}
\subsection{PtWSSe-IC4-S4}
\label{susec:144}
\begin{Verbatim}
1T-PtSSe/2H-WSSe IC4-S4                       
   1.00000000000000     
     3.4084401120541643    0.0000000000000000    0.0000000000000000
    -1.7042200560270822    2.9517957243980950    0.0000000000000000
     0.0000000000000000    0.0000000000000000   30.0000000000000000
   Se   S    Pt   W    Se   S 
     1     1     1     1     1     1
Direct
  0.3333333429999996  0.6666666870000029  0.4495298680933786
  0.6666666870000029  0.3333333429999996  0.5440850121014691
  0.0000000000000000  0.0000000000000000  0.4999503401741876
  0.6666666870000029  0.3333333429999996  0.6934589637136028
  0.0000000000000000  0.0000000000000000  0.7487491790718650
  0.0000000000000000  0.0000000000000000  0.6441266678455762
 
  0.00000000E+00  0.00000000E+00  0.00000000E+00
  0.00000000E+00  0.00000000E+00  0.00000000E+00
  0.00000000E+00  0.00000000E+00  0.00000000E+00
  0.00000000E+00  0.00000000E+00  0.00000000E+00
  0.00000000E+00  0.00000000E+00  0.00000000E+00
  0.00000000E+00  0.00000000E+00  0.00000000E+00
\end{Verbatim}
\subsection{PtWSSe-IC4-S5}
\label{susec:145}
\begin{Verbatim}
1T-PtSSe/2H-WSSe IC4-S5                       
   1.00000000000000     
     3.4146893946378687    0.0000000000000000    0.0000000000000000
    -1.7073446973189343    2.9572077619063082    0.0000000000000000
     0.0000000000000000    0.0000000000000000   30.0000000000000000
   Se   S    Pt   W    Se   S 
     1     1     1     1     1     1
Direct
  0.3333333429999996  0.6666666870000029  0.4521738576716317
  0.6666666870000029  0.3333333429999996  0.5463395769031845
 -0.0000000000000000  0.0000000000000000  0.5025330086526540
 -0.0000000000000000  0.0000000000000000  0.6909295556871327
  0.3333333429999996  0.6666666870000029  0.7461751917300713
  0.3333333429999996  0.6666666870000029  0.6417488403553199
 
  0.00000000E+00  0.00000000E+00  0.00000000E+00
  0.00000000E+00  0.00000000E+00  0.00000000E+00
  0.00000000E+00  0.00000000E+00  0.00000000E+00
  0.00000000E+00  0.00000000E+00  0.00000000E+00
  0.00000000E+00  0.00000000E+00  0.00000000E+00
  0.00000000E+00  0.00000000E+00  0.00000000E+00
\end{Verbatim}
\subsection{PtWSTe-IC1-S1}
\label{subsec:211}
\begin{Verbatim}
1T-PtSSe/2H-WSTe IC1-S1                       
   1.00000000000000     
     3.4772818972392847    0.0000000000000000    0.0000000000000000
    -1.7386409481233638    3.0114144594162449    0.0000000000000000
     0.0000000000000000    0.0000000000000000   30.0000000000000000
   S    Se   Pt   W    Te   S 
     1     1     1     1     1     1
Direct
  0.3333333429999996  0.6666666870000029  0.4425204450113185
  0.6666666870000029  0.3333333429999996  0.5342654730845595
  0.0000000000000000  0.0000000000000000  0.4857101603048761
  0.0000000000000000  0.0000000000000000  0.7021357900892298
  0.6666666870000029  0.3333333429999996  0.7642917203945387
  0.6666666870000029  0.3333333429999996  0.6542764411155204
 
\end{Verbatim}
\subsection{PtWSTe-IC1-S2}
\label{subsec:212}
\begin{Verbatim}
1T-PtSSe/2h-WSTe IC1-S2                       
   1.00000000000000     
     3.4833475634044397    0.0000000000000000    0.0000000000000000
    -1.7416737812054479    3.0166674803700588    0.0000000000000000
     0.0000000000000000    0.0000000000000000   30.0000000000000000
   S    Se   Pt   W    Te   S 
     1     1     1     1     1     1
Direct
  0.3333333429999996  0.6666666870000029  0.4532804024149755
  0.6666666870000029  0.3333333429999996  0.5445658978748824
  0.0000000000000000  0.0000000000000000  0.4964806709104863
  0.6666666870000029  0.3333333429999996  0.6915366505253999
  0.3333333429999996  0.6666666870000029  0.7535699716265043
  0.3333333429999996  0.6666666870000029  0.6437664366477804
 
  0.00000000E+00  0.00000000E+00  0.00000000E+00
  0.00000000E+00  0.00000000E+00  0.00000000E+00
  0.00000000E+00  0.00000000E+00  0.00000000E+00
  0.00000000E+00  0.00000000E+00  0.00000000E+00
  0.00000000E+00  0.00000000E+00  0.00000000E+00
  0.00000000E+00  0.00000000E+00  0.00000000E+00
\end{Verbatim}
\subsection{PtWSTe-IC1-S3}
\label{subsec:213}
\begin{Verbatim}
1T-PtSSe/2h-WSTe IC1-S3                       
   1.00000000000000     
     3.4824576217890830    0.0000000000000000    0.0000000000000000
    -1.7412288103994340    3.0158967683604110    0.0000000000000000
     0.0000000000000000    0.0000000000000000   30.0000000000000000
   S    Se   Pt   W    Te   S 
     1     1     1     1     1     1
Direct
  0.3333333429999996  0.6666666870000029  0.4539484070685873
  0.6666666870000029  0.3333333429999996  0.5453187177205834
  0.0000000000000000  0.0000000000000000  0.4971022069091120
  0.3333333429999996  0.6666666870000029  0.6908408068055536
  0.0000000000000000  0.0000000000000000  0.7529022111062318
  0.0000000000000000  0.0000000000000000  0.6430876803899537
 
  0.00000000E+00  0.00000000E+00  0.00000000E+00
  0.00000000E+00  0.00000000E+00  0.00000000E+00
  0.00000000E+00  0.00000000E+00  0.00000000E+00
  0.00000000E+00  0.00000000E+00  0.00000000E+00
  0.00000000E+00  0.00000000E+00  0.00000000E+00
  0.00000000E+00  0.00000000E+00  0.00000000E+00
\end{Verbatim}
\subsection{PtWSTe-IC1-S4}
\label{subsec:214}
\begin{Verbatim}
1T-PtSSe/2h-WSTe IC1-S4                       
   1.00000000000000     
     3.4794644876356431    0.0000000000000000    0.0000000000000000
    -1.7397322433215150    3.0133046381471482    0.0000000000000000
     0.0000000000000000    0.0000000000000000   30.0000000000000000
   S    Se   Pt   W    Te   S 
     1     1     1     1     1     1
Direct
  0.3333333429999996  0.6666666870000029  0.4519406391306546
  0.6666666870000029  0.3333333429999996  0.5434802802980130
  0.0000000000000000  0.0000000000000000  0.4951514852546737
  0.6666666870000029  0.3333333429999996  0.6927901465964510
  0.0000000000000000  0.0000000000000000  0.7548786953850453
  0.0000000000000000  0.0000000000000000  0.6449587833351984
 
  0.00000000E+00  0.00000000E+00  0.00000000E+00
  0.00000000E+00  0.00000000E+00  0.00000000E+00
  0.00000000E+00  0.00000000E+00  0.00000000E+00
  0.00000000E+00  0.00000000E+00  0.00000000E+00
  0.00000000E+00  0.00000000E+00  0.00000000E+00
  0.00000000E+00  0.00000000E+00  0.00000000E+00
\end{Verbatim}
\subsection{PtWSTe-IC1-S5}
\label{subsec:215}
\begin{Verbatim}
1T-PtSSe/2h-WSTe IC1-S5                       
   1.00000000000000     
     3.4863995188813885    0.0000000000000000    0.0000000000000000
    -1.7431997589435149    3.0193105513823477    0.0000000000000000
     0.0000000000000000    0.0000000000000000   30.0000000000000000
   S    Se   Pt   W    Te   S 
     1     1     1     1     1     1
Direct
  0.3333333429999996  0.6666666870000029  0.4542096339189783
  0.6666666870000029  0.3333333429999996  0.5453369642523356
  0.0000000000000000  0.0000000000000000  0.4973987170299239
  0.0000000000000000  0.0000000000000000  0.6906371324316254
  0.3333333429999996  0.6666666870000029  0.7526615160309404
  0.3333333429999996  0.6666666870000029  0.6429560663361968
 
  0.00000000E+00  0.00000000E+00  0.00000000E+00
  0.00000000E+00  0.00000000E+00  0.00000000E+00
  0.00000000E+00  0.00000000E+00  0.00000000E+00
  0.00000000E+00  0.00000000E+00  0.00000000E+00
  0.00000000E+00  0.00000000E+00  0.00000000E+00
  0.00000000E+00  0.00000000E+00  0.00000000E+00
\end{Verbatim}
\subsection{PtWSTe-IC2-S1}
\label{subsec:221}
\begin{Verbatim}
1T-PtSSe/2h-WSTe IC2-S1                       
   1.00000000000000     
     3.4762779463349451    0.0000000000000000    0.0000000000000000
    -1.7381389726713317    3.0105450124292519    0.0000000000000000
     0.0000000000000000    0.0000000000000000   30.0000000000000000
   S    Se   Pt   W    S    Te
     1     1     1     1     1     1
Direct
  0.3333333429999996  0.6666666870000029  0.4349783609550855
  0.6666666870000029  0.3333333429999996  0.5267713944190220
  0.0000000000000000  0.0000000000000000  0.4781776210463988
  0.0000000000000000  0.0000000000000000  0.7191389449460388
  0.6666666870000029  0.3333333429999996  0.7671407354984510
  0.6666666870000029  0.3333333429999996  0.6569929731349973
 
  0.00000000E+00  0.00000000E+00  0.00000000E+00
  0.00000000E+00  0.00000000E+00  0.00000000E+00
  0.00000000E+00  0.00000000E+00  0.00000000E+00
  0.00000000E+00  0.00000000E+00  0.00000000E+00
  0.00000000E+00  0.00000000E+00  0.00000000E+00
  0.00000000E+00  0.00000000E+00  0.00000000E+00
\end{Verbatim}
\subsection{PtWSTe-IC2-S2}
\label{subsec:222}
\begin{Verbatim}
1T-PtSSe/2h-WSTe IC2-S2                        
   1.00000000000000     
     3.4831991434342489    0.0000000000000000    0.0000000000000000
    -1.7415995712204304    3.0165389449053244    0.0000000000000000
     0.0000000000000000    0.0000000000000000   30.0000000000000000
   S    Se   Pt   W    S    Te
     1     1     1     1     1     1
Direct
  0.3333333429999996  0.6666666870000029  0.4454918352957229
  0.6666666870000029  0.3333333429999996  0.5368015062898726
 -0.0000000000000000 -0.0000000000000000  0.4887298658714790
  0.6666666870000029  0.3333333429999996  0.7087556922983305
  0.3333333429999996  0.6666666870000029  0.7566732864756478
  0.3333333429999996  0.6666666870000029  0.6467478437689833
 
  0.00000000E+00  0.00000000E+00  0.00000000E+00
  0.00000000E+00  0.00000000E+00  0.00000000E+00
  0.00000000E+00  0.00000000E+00  0.00000000E+00
  0.00000000E+00  0.00000000E+00  0.00000000E+00
  0.00000000E+00  0.00000000E+00  0.00000000E+00
  0.00000000E+00  0.00000000E+00  0.00000000E+00
\end{Verbatim}
\subsection{PtWSTe-IC2-S3}
\label{subsec:223}
\begin{Verbatim}
1T-PtSSe/2h-WSTe IC2-S3                       
   1.00000000000000     
     3.4818968629701734    0.0000000000000000    0.0000000000000000
    -1.7409484309897827    3.0154111369787300    0.0000000000000000
     0.0000000000000000    0.0000000000000000   30.0000000000000000
   S    Se   Pt   W    S    Te
     1     1     1     1     1     1
Direct
  0.3333333429999996  0.6666666870000029  0.4451048809763307
  0.6666666870000029  0.3333333429999996  0.5365687082109574
  0.0000000000000000  0.0000000000000000  0.4882305224719161
  0.3333333429999996  0.6666666870000029  0.7091065644817576
  0.0000000000000000  0.0000000000000000  0.7570632075440713
  0.0000000000000000  0.0000000000000000  0.6471261463149744
 
  0.00000000E+00  0.00000000E+00  0.00000000E+00
  0.00000000E+00  0.00000000E+00  0.00000000E+00
  0.00000000E+00  0.00000000E+00  0.00000000E+00
  0.00000000E+00  0.00000000E+00  0.00000000E+00
  0.00000000E+00  0.00000000E+00  0.00000000E+00
  0.00000000E+00  0.00000000E+00  0.00000000E+00
\end{Verbatim}
\subsection{PtWSTe-IC2-S4}
\label{subsec:224}
\begin{Verbatim}
1T-PtSSe/2h-WSTe IC2-S4                       
   1.00000000000000     
     3.4795080646657670    0.0000000000000000    0.0000000000000000
    -1.7397540318358087    3.0133423769827137    0.0000000000000000
     0.0000000000000000    0.0000000000000000   30.0000000000000000
   S    Se   Pt   W    S    Te
     1     1     1     1     1     1
Direct
  0.3333333429999996  0.6666666870000029  0.4456260285123648
  0.6666666870000029  0.3333333429999996  0.5371525754936428
  0.0000000000000000  0.0000000000000000  0.4888448875107798
  0.6666666870000029  0.3333333429999996  0.7085659489249849
  0.0000000000000000  0.0000000000000000  0.7565407985230550
  0.0000000000000000  0.0000000000000000  0.6464697910352015
 
  0.00000000E+00  0.00000000E+00  0.00000000E+00
  0.00000000E+00  0.00000000E+00  0.00000000E+00
  0.00000000E+00  0.00000000E+00  0.00000000E+00
  0.00000000E+00  0.00000000E+00  0.00000000E+00
  0.00000000E+00  0.00000000E+00  0.00000000E+00
  0.00000000E+00  0.00000000E+00  0.00000000E+00
\end{Verbatim}
\subsection{PtWSTe-IC2-S5}
\label{subsec:225}
\begin{Verbatim}
1T-PtSSe/2h-WSTe IC2-S5                       
   1.00000000000000     
     3.4825355069032131    0.0000000000000000    0.0000000000000000
    -1.7412677529548022    3.0159642188484144    0.0000000000000000
     0.0000000000000000    0.0000000000000000   30.0000000000000000
   S    Se   Pt   W    S    Te
     1     1     1     1     1     1
Direct
  0.3333333429999996  0.6666666870000029  0.4430371341384287
  0.6666666870000029  0.3333333429999996  0.5344469520433890
  0.0000000000000000  0.0000000000000000  0.4862508796405720
  0.0000000000000000  0.0000000000000000  0.7111682551437895
  0.3333333429999996  0.6666666870000029  0.7591151417206916
  0.3333333429999996  0.6666666870000029  0.6491816673132078
 
  0.00000000E+00  0.00000000E+00  0.00000000E+00
  0.00000000E+00  0.00000000E+00  0.00000000E+00
  0.00000000E+00  0.00000000E+00  0.00000000E+00
  0.00000000E+00  0.00000000E+00  0.00000000E+00
  0.00000000E+00  0.00000000E+00  0.00000000E+00
  0.00000000E+00  0.00000000E+00  0.00000000E+00
\end{Verbatim}
\subsection{PtWSTe-IC3-S1}
\label{subsec:231}
\begin{Verbatim}
1T-PtSSe/2h-WSTe IC3-S1                       
   1.00000000000000     
     3.4773139820300880    0.0000000000000000    0.0000000000000000
    -1.7386569905166434    3.0114422456641714    0.0000000000000000
     0.0000000000000000    0.0000000000000000   30.0000000000000000
   Se   S    Pt   W    S    Te
     1     1     1     1     1     1
Direct
  0.3333333429999996  0.6666666870000029  0.4358463877343342
  0.6666666870000029  0.3333333429999996  0.5275814340316172
 -0.0000000000000000  0.0000000000000000  0.4845197289728662
 -0.0000000000000000 -0.0000000000000000  0.7165312236894462
  0.6666666870000029  0.3333333429999996  0.7644121896151053
  0.6666666870000029  0.3333333429999996  0.6543090659566667
 
  0.00000000E+00  0.00000000E+00  0.00000000E+00
  0.00000000E+00  0.00000000E+00  0.00000000E+00
  0.00000000E+00  0.00000000E+00  0.00000000E+00
  0.00000000E+00  0.00000000E+00  0.00000000E+00
  0.00000000E+00  0.00000000E+00  0.00000000E+00
  0.00000000E+00  0.00000000E+00  0.00000000E+00
\end{Verbatim}
\subsection{PtWSTe-IC3-S2}
\label{subsec:232}
\begin{Verbatim}
1T-PtSSe/2h-WSTe IC3-S2                       
   1.00000000000000     
     3.4840224070527088    0.0000000000000000    0.0000000000000000
    -1.7420112030294430    3.0172519121119334    0.0000000000000000
     0.0000000000000000    0.0000000000000000   30.0000000000000000
   Se   S    Pt   W    S    Te
     1     1     1     1     1     1
Direct
  0.3333333429999996  0.6666666870000029  0.4494895844241584
  0.6666666870000029  0.3333333429999996  0.5407649089211455
  0.0000000000000000 -0.0000000000000000  0.4982023342872547
  0.6666666870000029  0.3333333429999996  0.7029723698235643
  0.3333333429999996  0.6666666870000029  0.7508236656838581
  0.3333333429999996  0.6666666870000029  0.6409471668600337
 
  0.00000000E+00  0.00000000E+00  0.00000000E+00
  0.00000000E+00  0.00000000E+00  0.00000000E+00
  0.00000000E+00  0.00000000E+00  0.00000000E+00
  0.00000000E+00  0.00000000E+00  0.00000000E+00
  0.00000000E+00  0.00000000E+00  0.00000000E+00
  0.00000000E+00  0.00000000E+00  0.00000000E+00
\end{Verbatim}
\subsection{PtWSTe-IC3-S3}
\label{subsec:233}
\begin{Verbatim}
1T-PtSSe/2h-WSTe IC3-S3                       
   1.00000000000000     
     3.4829213461749169    0.0000000000000000    0.0000000000000000
    -1.7414606725914161    3.0162983654587494    0.0000000000000000
     0.0000000000000000    0.0000000000000000   30.0000000000000000
   Se   S    Pt   W    S    Te
     1     1     1     1     1     1
Direct
  0.3333333429999996  0.6666666870000029  0.4470184496378175
  0.6666666870000029  0.3333333429999996  0.5385037520881745
 -0.0000000000000000  0.0000000000000000  0.4956377094661331
  0.3333333429999996  0.6666666870000029  0.7054465295455993
 -0.0000000000000000 -0.0000000000000000  0.7532237642420030
 -0.0000000000000000 -0.0000000000000000  0.6433698250203016
 
  0.00000000E+00  0.00000000E+00  0.00000000E+00
  0.00000000E+00  0.00000000E+00  0.00000000E+00
  0.00000000E+00  0.00000000E+00  0.00000000E+00
  0.00000000E+00  0.00000000E+00  0.00000000E+00
  0.00000000E+00  0.00000000E+00  0.00000000E+00
  0.00000000E+00  0.00000000E+00  0.00000000E+00
\end{Verbatim}
\subsection{PtWSTe-IC3-S4}
\label{subsec:234}
\begin{Verbatim}
1T-PtSSe/2h-WSTe IC4-S4                       
   1.00000000000000     
     3.4788774766185937    0.0000000000000000    0.0000000000000000
    -1.7394387378136809    3.0127962715536589    0.0000000000000000
     0.0000000000000000    0.0000000000000000   30.0000000000000000
   Se   S    Pt   W    S    Te
     1     1     1     1     1     1
Direct
  0.3333333429999996  0.6666666870000029  0.4473738191102769
  0.6666666870000029  0.3333333429999996  0.5389949350190538
  0.0000000000000000  0.0000000000000000  0.4960996888028775
  0.6666666870000029  0.3333333429999996  0.7050405280673608
 -0.0000000000000000  0.0000000000000000  0.7528740145788501
 -0.0000000000000000 -0.0000000000000000  0.6428170444216099
 
  0.00000000E+00  0.00000000E+00  0.00000000E+00
  0.00000000E+00  0.00000000E+00  0.00000000E+00
  0.00000000E+00  0.00000000E+00  0.00000000E+00
  0.00000000E+00  0.00000000E+00  0.00000000E+00
  0.00000000E+00  0.00000000E+00  0.00000000E+00
  0.00000000E+00  0.00000000E+00  0.00000000E+00
\end{Verbatim}
\subsection{PtWSTe-IC3-S5}
\label{subsec:235}
\begin{Verbatim}
1T-PtSSe/2h-WSTe IC3-S5                       
   1.00000000000000     
     3.4852304492775472    0.0000000000000000    0.0000000000000000
    -1.7426152241418515    3.0182981074031483    0.0000000000000000
     0.0000000000000000    0.0000000000000000   30.0000000000000000
   Se   S    Pt   W    S    Te
     1     1     1     1     1     1
Direct
  0.3333333429999996  0.6666666870000029  0.4466281759440420
  0.6666666870000029  0.3333333429999996  0.5378976933167934
  0.0000000000000000  0.0000000000000000  0.4952773236968255
  0.0000000000000000  0.0000000000000000  0.7058480116748171
  0.3333333429999996  0.6666666870000029  0.7536845915406190
  0.3333333429999996  0.6666666870000029  0.6438642338269176
 
  0.00000000E+00  0.00000000E+00  0.00000000E+00
  0.00000000E+00  0.00000000E+00  0.00000000E+00
  0.00000000E+00  0.00000000E+00  0.00000000E+00
  0.00000000E+00  0.00000000E+00  0.00000000E+00
  0.00000000E+00  0.00000000E+00  0.00000000E+00
  0.00000000E+00  0.00000000E+00  0.00000000E+00
\end{Verbatim}
\subsection{PtWSTe-IC4-S1}
\label{subsec:241}
\begin{Verbatim}
1T-PtSSe/2h-WSTe IC4-S1                       
   1.00000000000000     
     3.4779054209918110    0.0000000000000000    0.0000000000000000
    -1.7389527100031243    3.0119544467855177    0.0000000000000000
     0.0000000000000000    0.0000000000000000   30.0000000000000000
   Se   S    Pt   W    Te   S 
     1     1     1     1     1     1
Direct
  0.3333333429999996  0.6666666870000029  0.4424856921252101
  0.6666666870000029  0.3333333429999996  0.5341417315414176
  0.0000000000000000  0.0000000000000000  0.4911316111622241
  0.0000000000000000  0.0000000000000000  0.7003951248373426
  0.6666666870000029  0.3333333429999996  0.7625292441256590
  0.6666666870000029  0.3333333429999996  0.6525166262081896
 
  0.00000000E+00  0.00000000E+00  0.00000000E+00
  0.00000000E+00  0.00000000E+00  0.00000000E+00
  0.00000000E+00  0.00000000E+00  0.00000000E+00
  0.00000000E+00  0.00000000E+00  0.00000000E+00
  0.00000000E+00  0.00000000E+00  0.00000000E+00
  0.00000000E+00  0.00000000E+00  0.00000000E+00
\end{Verbatim}
\subsection{PtWSTe-IC4-S2}
\label{subsec:242}
\begin{Verbatim}
1T-PtSSe/2h-WSTe IC4-S2                       
   1.00000000000000     
     3.4821621231468138    0.0000000000000000    0.0000000000000000
    -1.7410810610766745    3.0156408589921595    0.0000000000000000
     0.0000000000000000    0.0000000000000000   30.0000000000000000
   Se   S    Pt   W    Te   S 
     1     1     1     1     1     1
Direct
  0.3333333429999996  0.6666666870000029  0.4543282024769582
  0.6666666870000029  0.3333333429999996  0.5456382430891580
  0.0000000000000000  0.0000000000000000  0.5029788235505137
  0.6666666870000029  0.3333333429999996  0.6886771211777258
  0.3333333429999996  0.6666666870000029  0.7507038424714949
  0.3333333429999996  0.6666666870000029  0.6408737972341711
 
  0.00000000E+00  0.00000000E+00  0.00000000E+00
  0.00000000E+00  0.00000000E+00  0.00000000E+00
  0.00000000E+00  0.00000000E+00  0.00000000E+00
  0.00000000E+00  0.00000000E+00  0.00000000E+00
  0.00000000E+00  0.00000000E+00  0.00000000E+00
  0.00000000E+00  0.00000000E+00  0.00000000E+00
\end{Verbatim}
\subsection{PtWSTe-IC4-S3}
\label{subsec:243}
\begin{Verbatim}
1T-PtSSe/2h-WSTe IC4-S3                       
   1.00000000000000     
     3.4817214545571811    0.0000000000000000    0.0000000000000000
    -1.7408607267860070    3.0152592288372477    0.0000000000000000
     0.0000000000000000    0.0000000000000000   30.0000000000000000
   Se   S    Pt   W    Te   S 
     1     1     1     1     1     1
Direct
  0.3333333429999996  0.6666666870000029  0.4544632954207870
  0.6666666870000029  0.3333333429999996  0.5458480215196317
  0.0000000000000000  0.0000000000000000  0.5030764265372980
  0.3333333429999996  0.6666666870000029  0.6885014866547579
  0.0000000000000000  0.0000000000000000  0.7505663679230210
  0.0000000000000000  0.0000000000000000  0.6407444319444906
 
  0.00000000E+00  0.00000000E+00  0.00000000E+00
  0.00000000E+00  0.00000000E+00  0.00000000E+00
  0.00000000E+00  0.00000000E+00  0.00000000E+00
  0.00000000E+00  0.00000000E+00  0.00000000E+00
  0.00000000E+00  0.00000000E+00  0.00000000E+00
  0.00000000E+00  0.00000000E+00  0.00000000E+00
\end{Verbatim}
\subsection{PtWSTe-IC4-S4}
\label{subsec:244}
\begin{Verbatim}
1T-PtSSe/2h-WSTe IC4-S4                       
   1.00000000000000     
     3.4790919718253548    0.0000000000000000    0.0000000000000000
    -1.7395459854164070    3.0129820299919965    0.0000000000000000
     0.0000000000000000    0.0000000000000000   30.0000000000000000
   S    Se   Pt   W    S    Te
     1     1     1     1     1     1
Direct
  0.3333333429999996  0.6666666870000029  0.4521337198636459
  0.6666666870000029  0.3333333429999996  0.5436528021365987
  0.0000000000000000  0.0000000000000000  0.5007949744709634
  0.6666666870000029  0.3333333429999996  0.6907972675946752
  0.0000000000000000  0.0000000000000000  0.7528780702314322
  0.0000000000000000  0.0000000000000000  0.6429431957027560
 
  0.00000000E+00  0.00000000E+00  0.00000000E+00
  0.00000000E+00  0.00000000E+00  0.00000000E+00
  0.00000000E+00  0.00000000E+00  0.00000000E+00
  0.00000000E+00  0.00000000E+00  0.00000000E+00
  0.00000000E+00  0.00000000E+00  0.00000000E+00
  0.00000000E+00  0.00000000E+00  0.00000000E+00
\end{Verbatim}
\subsection{PtWSTe-IC4-S5}
\label{subsec:245}
\begin{Verbatim}
1T-PtSSe/2h-WSTe IC4-S5                       
   1.00000000000000     
     3.4868002123798627    0.0000000000000000    0.0000000000000000
    -1.7434001056928561    3.0196575621315414    0.0000000000000000
     0.0000000000000000    0.0000000000000000   30.0000000000000000
   Se   S    Pt   W    Te   S 
     1     1     1     1     1     1
Direct
  0.3333333429999996  0.6666666870000029  0.4566697095048511
  0.6666666870000029  0.3333333429999996  0.5477069085015316
  0.0000000000000000  0.0000000000000000  0.5052738077656471
  0.0000000000000000  0.0000000000000000  0.6864011035002306
  0.3333333429999996  0.6666666870000029  0.7484023394658408
  0.3333333429999996  0.6666666870000029  0.6387461612618992
 
  0.00000000E+00  0.00000000E+00  0.00000000E+00
  0.00000000E+00  0.00000000E+00  0.00000000E+00
  0.00000000E+00  0.00000000E+00  0.00000000E+00
  0.00000000E+00  0.00000000E+00  0.00000000E+00
  0.00000000E+00  0.00000000E+00  0.00000000E+00
  0.00000000E+00  0.00000000E+00  0.00000000E+00
\end{Verbatim}
\subsection{PtWSeTe-IC1-S1}
\label{subsec:311}
\begin{Verbatim}
1T-PtSSe/2H-WSeTe IC1-S1                      
   1.00000000000000     
     3.5168703876598477    0.0000000000000000    0.0000000000000000
    -1.7584351943262397    3.0456990978764451    0.0000000000000000
     0.0000000000000000    0.0000000000000000   30.0000000000000000
   S    Se   Pt   W    Te   Se
     1     1     1     1     1     1
Direct
  0.3333333429999996  0.6666666870000029  0.4418875007299832
  0.6666666870000029  0.3333333429999996  0.5319936086147834
  0.0000000000000000  0.0000000000000000  0.4843878911537374
  0.0000000000000000  0.0000000000000000  0.7081077419176012
  0.6666666870000029  0.3333333429999996  0.7693566075967624
  0.6666666870000029  0.3333333429999996  0.6550532589871523

\end{Verbatim}
\subsection{PtWSeTe-IC1-S2}
\label{subsec:312}
\begin{Verbatim}
1T-PtSSe/2H-WSeTe IC1-S2                       
   1.00000000000000     
     3.5244264124533329    0.0000000000000000    0.0000000000000000
    -1.7622132067239247    3.0522428072988412    0.0000000000000000
     0.0000000000000000    0.0000000000000000   30.0000000000000000
   S    Se   Pt   W    Te   Se
     1     1     1     1     1     1
Direct
  0.3333333429999996  0.6666666870000029  0.4538373080794429
  0.6666666870000029  0.3333333429999996  0.5434247514052615
  0.0000000000000000  0.0000000000000000  0.4963357210069148
  0.6666666870000029  0.3333333429999996  0.6963326280162434
  0.3333333429999996  0.6666666870000029  0.7574448822896187
  0.3333333429999996  0.6666666870000029  0.6434113182025101

\end{Verbatim}
\subsection{PtWSeTe-IC1-S3}
\label{subsec:313}
\begin{Verbatim}
1T-PtSSe/2H-WSeTe IC1-S3
   1.00000000000000     
     3.5227995018906313    0.0000000000000000    0.0000000000000000
    -1.7613997514424606    3.0508338614226922    0.0000000000000000
     0.0000000000000000    0.0000000000000000   30.0000000000000000
   S    Se   Pt   W    Te   Se
     1     1     1     1     1     1
Direct
  0.3333333429999996  0.6666666870000029  0.4535561279676372
  0.6666666870000029  0.3333333429999996  0.5432880871346057
  0.0000000000000000  0.0000000000000000  0.4959960902790002
  0.3333333429999996  0.6666666870000029  0.6965663643290867
  0.0000000000000000  0.0000000000000000  0.7577253175838550
  0.0000000000000000  0.0000000000000000  0.6436546217058421
 
  0.00000000E+00  0.00000000E+00  0.00000000E+00
  0.00000000E+00  0.00000000E+00  0.00000000E+00
  0.00000000E+00  0.00000000E+00  0.00000000E+00
  0.00000000E+00  0.00000000E+00  0.00000000E+00
  0.00000000E+00  0.00000000E+00  0.00000000E+00
  0.00000000E+00  0.00000000E+00  0.00000000E+00
\end{Verbatim}
\subsection{PtWSeTe-IC1-S4}
\label{subsec:314}
\begin{Verbatim}
1T-PtSSe/2H-WSeTe IC1-S4                      
   1.00000000000000     
     3.5194436032417635    0.0000000000000000    0.0000000000000000
    -1.7597218021179968    3.0479275679466693    0.0000000000000000
     0.0000000000000000    0.0000000000000000   30.0000000000000000
   S    Se   Pt   W    Te   Se
     1     1     1     1     1     1
Direct
  0.3333333429999996  0.6666666870000029  0.4520073045373984
  0.6666666870000029  0.3333333429999996  0.5418958272227055
  0.0000000000000000  0.0000000000000000  0.4945261279591894
  0.6666666870000029  0.3333333429999996  0.6980604427328313
  0.0000000000000000  0.0000000000000000  0.7592480636100417
  0.0000000000000000  0.0000000000000000  0.6450488429377966
 
  0.00000000E+00  0.00000000E+00  0.00000000E+00
  0.00000000E+00  0.00000000E+00  0.00000000E+00
  0.00000000E+00  0.00000000E+00  0.00000000E+00
  0.00000000E+00  0.00000000E+00  0.00000000E+00
  0.00000000E+00  0.00000000E+00  0.00000000E+00
  0.00000000E+00  0.00000000E+00  0.00000000E+00
\end{Verbatim}
\subsection{PtWSeTe-IC1-S5}
\label{subsec:315}
\begin{Verbatim}
1T-PtSSe/2H-WSeTe IC1-S5                       
   1.00000000000000     
     3.5267972790276216    0.0000000000000000    0.0000000000000000
    -1.7633986400113248    3.0542960379809130    0.0000000000000000
     0.0000000000000000    0.0000000000000000   30.0000000000000000
   S    Se   Pt   W    Te   Se
     1     1     1     1     1     1
Direct
  0.3333333429999996  0.6666666870000029  0.4536719051750424
  0.6666666870000029  0.3333333429999996  0.5431669528554366
  0.0000000000000000  0.0000000000000000  0.4961623275542237
  0.0000000000000000  0.0000000000000000  0.6965014721581895
  0.3333333429999996  0.6666666870000029  0.7576199081302093
  0.3333333429999996  0.6666666870000029  0.6436640431269183
 
  0.00000000E+00  0.00000000E+00  0.00000000E+00
  0.00000000E+00  0.00000000E+00  0.00000000E+00
  0.00000000E+00  0.00000000E+00  0.00000000E+00
  0.00000000E+00  0.00000000E+00  0.00000000E+00
  0.00000000E+00  0.00000000E+00  0.00000000E+00
  0.00000000E+00  0.00000000E+00  0.00000000E+00
\end{Verbatim}
\subsection{PtWSeTe-IC2-S1}
\label{subsec:321}
\begin{Verbatim}
1T-PtSSe/2H-WSeTe IC2-S1                      
   1.00000000000000     
     3.5160234608445471    0.0000000000000000    0.0000000000000000
    -1.7580117309181189    3.0449656377339207    0.0000000000000000
     0.0000000000000000    0.0000000000000000   30.0000000000000000
   S    Se   Pt   W    Se   Te
     1     1     1     1     1     1
Direct
  0.3333333429999996  0.6666666870000029  0.4369923468089638
  0.6666666870000029  0.3333333429999996  0.5271406521424851
  0.0000000000000000  0.0000000000000000  0.4795035852566656
  0.0000000000000000  0.0000000000000000  0.7184123245369349
  0.6666666870000029  0.3333333429999996  0.7715640384154696
  0.6666666870000029  0.3333333429999996  0.6571736618395363
 
  0.00000000E+00  0.00000000E+00  0.00000000E+00
  0.00000000E+00  0.00000000E+00  0.00000000E+00
  0.00000000E+00  0.00000000E+00  0.00000000E+00
  0.00000000E+00  0.00000000E+00  0.00000000E+00
  0.00000000E+00  0.00000000E+00  0.00000000E+00
  0.00000000E+00  0.00000000E+00  0.00000000E+00
\end{Verbatim}
\subsection{PtWSeTe-IC2-S2}
\label{subsec:322}
\begin{Verbatim}
1T-PtSSe/2H-WSeTe IC2-S2                      
   1.00000000000000     
     3.5238223771283343    0.0000000000000000    0.0000000000000000
    -1.7619111890614558    3.0517196973622212    0.0000000000000000
     0.0000000000000000    0.0000000000000000   30.0000000000000000
   S    Se   Pt   W    Se   Te
     1     1     1     1     1     1
Direct
  0.3333333429999996  0.6666666870000029  0.4485520967204479
  0.6666666870000029  0.3333333429999996  0.5381821110632359
  0.0000000000000000  0.0000000000000000  0.4910996084978834
  0.6666666870000029  0.3333333429999996  0.7069935142795600
  0.3333333429999996  0.6666666870000029  0.7600438888593004
  0.3333333429999996  0.6666666870000029  0.6459153895795495
 
  0.00000000E+00  0.00000000E+00  0.00000000E+00
  0.00000000E+00  0.00000000E+00  0.00000000E+00
  0.00000000E+00  0.00000000E+00  0.00000000E+00
  0.00000000E+00  0.00000000E+00  0.00000000E+00
  0.00000000E+00  0.00000000E+00  0.00000000E+00
  0.00000000E+00  0.00000000E+00  0.00000000E+00
\end{Verbatim}
\subsection{PtWSeTe-IC2-S3}
\label{subsec:323}
\begin{Verbatim}
1T-PtSSe/2H-WSeTe IC2-S3                      
   1.00000000000000     
     3.5221482033980629    0.0000000000000000    0.0000000000000000
    -1.7610741021958407    3.0502698203824035    0.0000000000000000
     0.0000000000000000    0.0000000000000000   30.0000000000000000
   S    Se   Pt   W    Se   Te
     1     1     1     1     1     1
Direct
  0.3333333429999996  0.6666666870000029  0.4479030176942871
  0.6666666870000029  0.3333333429999996  0.5377082755606040
  0.0000000000000000  0.0000000000000000  0.4903358239259106
  0.3333333429999996  0.6666666870000029  0.7076025193059436
  0.0000000000000000  0.0000000000000000  0.7606964435684915
  0.0000000000000000  0.0000000000000000  0.6465405289447830
 
  0.00000000E+00  0.00000000E+00  0.00000000E+00
  0.00000000E+00  0.00000000E+00  0.00000000E+00
  0.00000000E+00  0.00000000E+00  0.00000000E+00
  0.00000000E+00  0.00000000E+00  0.00000000E+00
  0.00000000E+00  0.00000000E+00  0.00000000E+00
  0.00000000E+00  0.00000000E+00  0.00000000E+00
\end{Verbatim}
\subsection{PtWSeTe-IC2-S4}
\label{subsec:324}
\VerbatimInput{POSCAR_PtWSeTe-IC2-S4}
\subsection{PtWSeTe-IC2-S5}
\label{subsec:325}
\VerbatimInput{POSCAR_PtWSeTe-IC2-S5}
\subsection{PtWSeTe-IC3-S1}
\label{subsec:331}
\VerbatimInput{POSCAR_PtWSeTe-IC3-S1}
\subsection{PtWSeTe-IC3-S2}
\label{subsec:332}
\VerbatimInput{POSCAR_PtWSeTe-IC3-S2}
\subsection{PtWSeTe-IC3-S3}
\label{subsec:333}
\VerbatimInput{POSCAR_PtWSeTe-IC3-S3}
\subsection{PtWSeTe-IC3-S4}
\label{subsec:334}
\VerbatimInput{POSCAR_PtWSeTe-IC3-S4}
\subsection{PtWSeTe-IC3-S5}
\label{subsec:335}
\VerbatimInput{POSCAR_PtWSeTe-IC3-S5}
\subsection{PtWSeTe-IC4-S1}
\label{subsec:341}
\VerbatimInput{POSCAR_PtWSeTe-IC4-S1}
\subsection{PtWSeTe-IC4-S2}
\label{subsec:342}
\VerbatimInput{POSCAR_PtWSeTe-IC4-S2}
\subsection{PtWSeTe-IC4-S3}
\label{subsec:343}
\VerbatimInput{POSCAR_PtWSeTe-IC4-S3}
\subsection{PtWSeTe-IC4-S4}
\label{subsec:344}
\VerbatimInput{POSCAR_PtWSeTe-IC4-S4}
\subsection{PtWSeTe-IC4-S5}
\label{subsec:345}
\VerbatimInput{POSCAR_PtWSeTe-IC4-S5}

\end{document}